\documentclass[11pt]{article}

\usepackage[final]{acl}

\usepackage{times}
\usepackage{latexsym}

\usepackage[T1]{fontenc}

\usepackage[utf8]{inputenc}

\usepackage{microtype}

\usepackage{inconsolata}

\usepackage{graphicx}

\usepackage{amssymb}
\usepackage{amsmath}
\usepackage{booktabs}
\usepackage{multirow}
\usepackage{graphicx}
\usepackage{booktabs}
\usepackage{multirow}
\usepackage{graphicx}
\usepackage[table,xcdraw]{xcolor}
\usepackage{algorithm}
\usepackage{algorithmic}

\usepackage{underscore}
\usepackage[normalem]{ulem}
\usepackage{enumitem}
\useunder{\uline}{\ul}{}

\title{SEE: Signal Embedding Energy for Quantifying Noise Interference \\ in Large Audio Language Models}

\author{
 \textbf{Yuanhe Zhang\textsuperscript{1,$^\star$}}, 
 \textbf{Jiayu Tian\textsuperscript{2,$^\star$}}, 
 \textbf{Yibo Zhang\textsuperscript{1}}, 
 \textbf{Shilinlu Yan\textsuperscript{1},}
 \\
 \textbf{Liang Lin\textsuperscript{3},}
 \textbf{Zhenhong Zhou\textsuperscript{4},} 
 \textbf{Li Sun\textsuperscript{1},} 
 \textbf{Sen Su\textsuperscript{1, 5, $^\dagger$}} 
\\ \textsuperscript{\rm 1}Beijing University of Posts and Telecommunications \quad
\textsuperscript{\rm 2}North China Electric Power University
\\ \textsuperscript{\rm 3}Institute of Information Engineering, Chinese Academy of Sciences
\\ \textsuperscript{\rm 4}Nanyang Technological University \quad
\textsuperscript{\rm 5}Chongqing University of Posts and Telecommunications
\\ \{charmes-zhang, zhangyibo2023, lulu\_land, lsun, susen\}@bupt.edu.cn;
\\ jiayu@ncepu.edu.cn; linliang@iie.ac.cn; zhenhong001@e.ntu.edu.sg
}

\begin{document}
\maketitle
\begingroup
\renewcommand\thefootnote{}\footnotemark
\footnotetext{$\star$ indicates equal contribution. $\dagger$ indicates corresponding author.}
\endgroup
\begin{abstract}
Large Audio Language Models (LALMs) have been widely applied in real-time scenarios, such as in-car assistants and online meeting comprehension.
In practice, audio inputs are often corrupted by device and environmental noise, leading to performance degradation.
However, existing LALM studies on noise lack quantitative analysis and rely mainly on intuition and empirical observation, thus failing to understand practical robustness.
To address this issue, we introduce \textbf{S}ignal \textbf{E}mbedding \textbf{E}nergy (\textbf{SEE}), a method for quantifying the impact of noise intensity on LALM inputs, enabling the differentiation of LALM robustness in real-world deployments.
SEE introduces a perspective based on structured activation subspaces derived from the model's internal representations, which more accurately captures its perception of noise than raw audio features.
Across experiments, SEE exhibits a strong correlation with LALM performance, achieving a correlation of 0.98.
Surprisingly, traditional audio denoising methods are only marginally effective for LALMs, and, in some cases, even increase SEE and impair performance.
This suggests a mismatch between speech-centric denoising objectives and the noise sensitivity of modern LALMs.
Therefore, we propose a mitigation strategy derived from SEE to denoise LALM inputs, outperforming existing denoising methods.
This paper introduces a novel metric for noise quantification in LALMs, providing guidance for robustness improvements in real-world deployments. 
Our code is publicly available at \url{https://github.com/jyutian/SEEN}.

\end{abstract}

\section{Introduction}
\begin{figure}[t]
    \centering
    \includegraphics[width=\columnwidth]{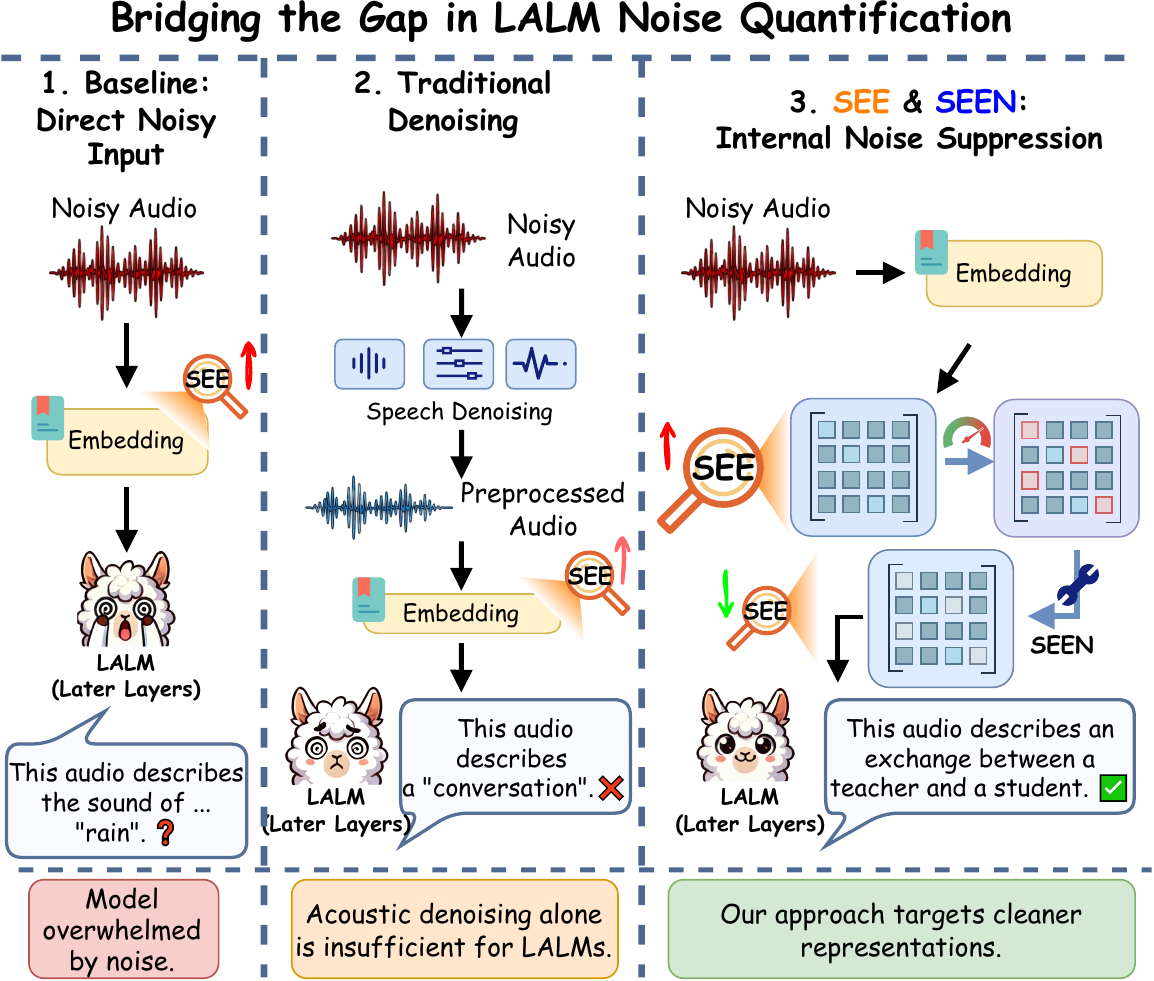}
    \caption{Motivation and overview of representation-level noise robustness in LALMs. Waveform-level denoising improves acoustic quality but may introduce semantic interference, which is quantified by Signal Embedding Energy (SEE) and mitigated by SEEN.}
    \label{fig:intro}
    \vspace{-9pt}
\end{figure}
\begin{figure*}[t]
    \centering
    \includegraphics[width=\textwidth]{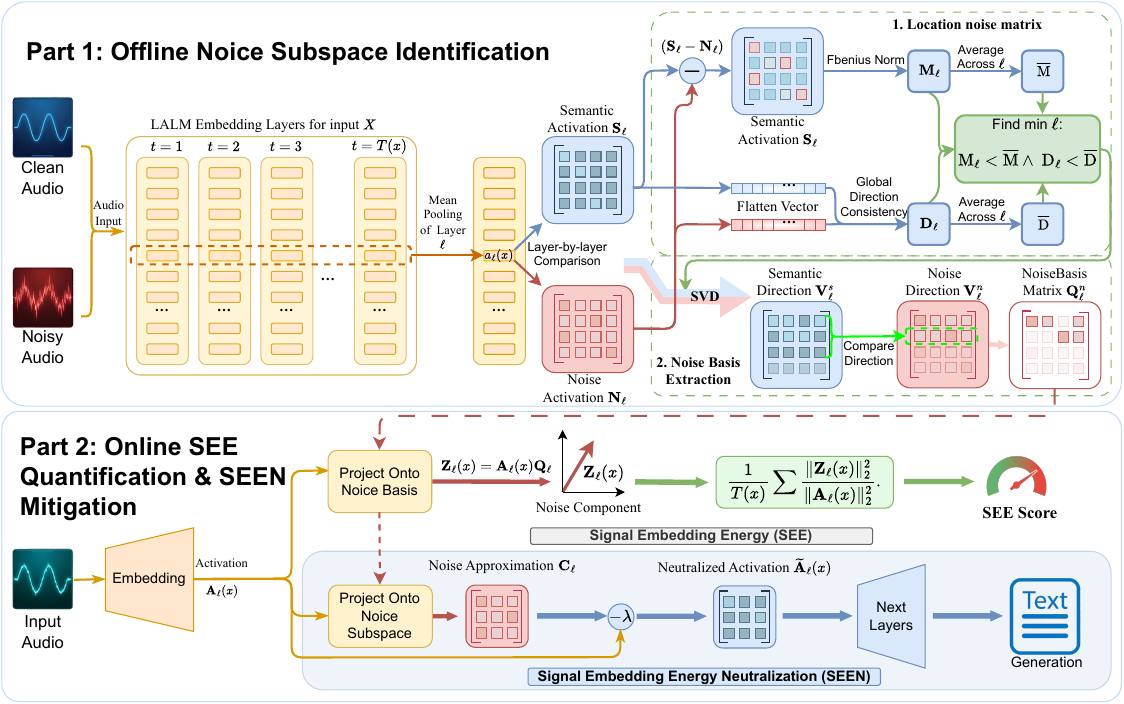}
    \caption{Offline, we construct noise activation matrices, localize noise dominant layers, and extract principal noise directions to form a noise subspace. Online, SEE quantifies activation energy projected onto this subspace, and SEEN mitigates noise by removing the projected components at the embedding layers.}
    \label{fig:main}
    \vspace{-9pt}
\end{figure*}
Large Audio Language Models (LALMs) are increasingly deployed in real-time applications, such as in-car assistants, where audio serves as the primary interface for language understanding and interaction~\cite{rubenstein2023audiopalm, radford2023robust,li2025applications}. In these real-world environments, device sampling imperfections and environmental noise corrupt the input waveform, turning a clean request into a signal with interference~\cite{gelbart2002double, deliyski2005adverse}. 
This corrupted signal is then fed into the LALMs, where the same noise that distorts the waveform also degrades the model's generation quality\cite{lin2025hidden,kumar2025robustness}. 
While quantifying this effect in LALMs remains challenging, existing studies typically rely on task performance~\cite{shen2024taskbench}, a proxy that requires large-scale evaluation and is difficult to leverage for mitigation.

To maximize performance on benchmarks, many LALMs are trained and validated under comparatively idealized conditions~\cite{chu2024qwen2,zhang2023speechgpt}. 
In real-world deployments, however, input audio is frequently corrupted by device artifacts and environmental noise, and the resulting impact on LALMs is still difficult to quantify in a principled way~\cite{li2025silence,kumar2025robustness,hou2025evaluating}. 
Although incorporating additional real-world data or applying speech enhancement may improve robustness, the lack of quantitative evaluation makes it challenging to assess these interventions and compare models reliably~\cite{goel2021robustness,hu2024large}. 
Existing studies often rely on outcome-level metrics such as performance on diverse tasks, which require large-scale benchmarking and provide limited guidance for diagnosing noise effects or informing mitigation~\cite{yang2024air,wang2025audiobench}. 
These limitations motivate a model-aware criterion that directly quantifies noise interference within LALMs.

To quantify semantic interference, we introduce \textbf{S}ignal \textbf{E}mbedding \textbf{E}nergy (\textbf{SEE}), a metric that measures noise intensity in the LALM embedding space.
SEE leverages the activation space of LALMs to disentangle the semantic and noise components of the input embedding. 
By analyzing directional and magnitude discrepancies across layers, we identify the primary points at which noise distorts semantic processing.
We then measure that captures the cumulative strength of noise activations, providing a quantification of semantic interference during generation.
Because SEE captures embedding-level inference bias, it serves as a direct proxy for the impact of noise on model generation.
Using SEE as a probe, we show that SEE increases monotonically with noise intensity and strongly correlates with generation quality, with a Pearson correlation~\cite{benesty2009pearson} coefficient of 0.98. 
Moreover, mainstream LALMs are vulnerable to real-world noise, and standard denoising pipelines often fail to reduce SEE and may even increase it, suggesting that acoustic denoising does not necessarily mitigate semantic interference.


Guided by SEE, we propose a defense that mitigates noise without any retraining. Concretely, we introduce \textbf{S}ignal \textbf{E}mbedding \textbf{E}nergy \textbf{N}eutralization (\textbf{SEEN}), which operates directly on the model's audio embeddings to minimize noise components.
Because SEEN targets embedding interference rather than waveform energy, it aligns the mitigation objective with SEE. In experiments, SEEN improves accuracy by 6.7\% over existing denoisings.

\noindent Our contributions are summarized as follows:
\begin{itemize}[noitemsep, topsep=0pt, parsep=0pt, leftmargin=*]
    \item We introduce \textbf{SEE}, a metric that quantifies the perturbation intensity in LALMs beyond traditional acoustic measures.
    \item We propose \textbf{SEEN}, a training‑free gating that optimizes audio embedding and reduces misjudgments under noise conditions.
    \item We conduct extensive experiments to analyze the risks of acoustic mitigation and evaluate the effectiveness of our study.
\end{itemize}

\section{Related Work}
\subsection{Large Audio Language Models}
LALMs extend the reasoning capabilities of Large Language Models(LLMs) to the auditory domain by integrating pre-trained audio encoders with textual backbones \cite{radford2023robust, chen2023beats}. Technically, while LLMs operate on discrete text tokens, LALMs typically align high-dimensional continuous signals with semantic spaces through discrete tokenization \cite{lakhotia2021generative, zhang2023speechgpt} or cross-modal adapters \cite{tang2023salmonn, chu2023qwen}. This architectural shift allows LALMs to capture paralinguistic cues and environmental contexts often inaccessible to text-only models, expanding application scenarios to sophisticated audio-centric reasoning. Representative frameworks include AudioPaLM \cite{rubenstein2023audiopalm}, Qwen-Audio \cite{Qwen2-Audio}, MiniCPM-o \cite{yao2024minicpm}, and StepAudio \cite{wu2025stepaudio2technicalreport}. Unlike cascaded pipelines, LALMs' end-to-end nature facilitates unified instruction following across diverse acoustic tasks \cite{gong2024listen, peng2024survey}. 

\subsection{Noise Robustness}
Traditional speech robustness primarily focuses on mitigating acoustic mismatch \cite{li2015robust} via multi-condition training, data augmentation \cite{ko2015audio, park2019specaugment}, or enhancement front-ends \cite{wang2018supervised} to minimize Word Error Rate \cite{watanabe2018espnet}. Conversely, the robustness of Large Audio Language Models (LALMs) centers on semantic reasoning stability and instruction-following integrity \cite{li2025isa, hou2025evaluating}. Given that non-stationary noise frequently induces "semantic hallucinations" \cite{hou2025evaluating}, research has pivoted from acoustic fidelity toward logical consistency and task reliability under environmental perturbations \cite{wang2025audiobench, xiong2025thinking, gopal2025explainable}.

\subsection{Speech Enhancement and Denoising Front-Ends}
The paradigm of speech enhancement has shifted from statistical signal processing to data-driven deep learning. Early classical methods, such as Wiener Filtering \cite{lim1978all}, Spectral Subtraction \cite{boll2003suppression}, and MMSE-STSA \cite{ephraim2003speech}, primarily targeted stationary noise through spectral estimation. Modern neural approaches have advanced this via spectral mapping \cite{xu2014regression}, end-to-end time-domain modeling like Conv-TasNet \cite{luo2019conv}, and generative frameworks including GANs \cite{pascual2017segan} and diffusion models \cite{lu2022conditional}. 
However, traditional Speech Enhancement prioritizes acoustic fidelity for task performance, often neglecting the semantic coherence required by LALMs. Consequently, denoising artifacts may improve signal metrics while undermining multimodal alignment and reasoning.

\section{Method}
In this section, we formally present a framework for analyzing and mitigating noise disturbances in LALMs at the level of internal embedding activations.
We first introduce Signal Embedding Energy (SEE), a metric that quantifies noise interference through semantic direction decomposition.
Then, we present Signal Embedding Energy Neutralization (SEEN), a training-free strategy that subtracts the noise components and improves downstream generation quality.

\subsection{Noise Substance Identification}
\noindent \textbf{Notation.}
We denote a LALM with $L$ observable activation function layer, indexed by $\ell\in\{1,\dots,L\}$.
For an audio input $ x \in X $, the activation at block $\ell$ is a matrix $\mathbf{A}_\ell(x)\in\mathbb{R}^{T(x)\times d_\ell}$, where $T(x)$ is the number of time steps and $d_\ell$ is the hidden width at layer $\ell$.

We use two aligned input sets, containing a semantic (clean) request set $X^s=\{x_1^s,x_2^s,\cdots,x_m^s\}$ and a pure noise set $X^n=\{x_1^n,x_2^n,\cdots,x_m^n\}$, where $m$ denotes the number of signal frames collected in the downstream application environment.

We keep dominant singular directions using threshold $\alpha$; treat a direction as ``noise-only'' if its cosine similarity to any dominant semantic direction is below $\delta=0.1$; and use $\varepsilon>0$ as a small constant for numerical stability.

\noindent \textbf{Separate the Noise Direction.}
Since audio inputs are represented as frame-level token sequences with different sequence lengths $T(x)$.
To balance the differences in sample lengths, we apply mean pooling over the time dimension at layer $\ell$:
\begin{equation}
\mathbf{a}_\ell(x)=\frac{1}{T(x)}\sum_{t=1}^{T(x)} \mathbf{A}_\ell(x)_{t,:}\in\mathbb{R}^{d_\ell},
\end{equation}
where $\mathbf{A}_\ell(x)_{t,:}$ denotes activation vector of the $t$-th frame token.

Next, we stack the pooled vectors across $M$ samples to form a joint representation space:
\begin{equation}
\begin{aligned}
\mathbf{S}_\ell=&
\begin{bmatrix}
\mathbf{a}_\ell(x_1^s) \quad
\cdots \quad
\mathbf{a}_\ell(x_M^s)
\end{bmatrix}^\top
\in\mathbb{R}^{M\times d_\ell}, \\
\mathbf{N}_\ell=&
\begin{bmatrix}
\mathbf{a}_\ell(x_1^n) \quad
\cdots \quad
\mathbf{a}_\ell(x_M^n)
\end{bmatrix}^\top
\in\mathbb{R}^{M\times d_\ell}.
\end{aligned}
\end{equation}
Here $\mathbf{S}_\ell$ is the semantic (clean) activation matrix and $\mathbf{N}_\ell$ is the noise activation matrix at layer $\ell$.

\noindent \textbf{Location Noise Matrix.}
To pinpoint the location where noise dominantly affects the encoding tendency, we characterize interference in two complementary aspects: magnitude and direction.

To identify where noise begins to alter semantic processing, we compute Frobenius Norm~\cite{bottcher2008frobenius} as the overall discrepancy energy using the difference matrix:
\begin{equation}
\begin{aligned}
\mathrm{M}_\ell=\sqrt{\sum_{i=1}^{M}\sum_{j=1}^{d_\ell}(\mathbf{S}_\ell(i,j)-\mathbf{N}_\ell(i,j))^2} \quad,
\end{aligned}
\end{equation}
here $\mathbf{S}_\ell(i,j)$ (resp. $\mathbf{N}_\ell(i,j)$) denotes the value of the pooled activation vector at feature dimension $j$ for the $i$-th clean (resp. noise) input at layer $\ell$.
We then use $\mathrm{vec}(\cdot)$ to flatten the matrix into a vector in $\mathbb{R}^{Md_\ell \times 1} $ and compute the global direction consistency:
\begin{equation}
\mathrm{D}_\ell=
\frac{\|\mathrm{vec}(\mathbf{S}_\ell)^\top\mathrm{vec}(\mathbf{N}_\ell)\|_2}
{\|\mathrm{vec}(\mathbf{S}_\ell)\|_2~\|\mathrm{vec}(\mathbf{N}_\ell)\|_2+\varepsilon},
\end{equation}
we use 
$\|\cdot\|_2$ to denote the Euclidean (L2) norm of a vector.
And its average value $\overline{\mathrm{M}}=\frac{1}{L}\sum_{\ell=1}^{L}\mathrm{M}_\ell$ and $\overline{\mathrm{D}}=\frac{1}{L}\sum_{\ell=1}^{L}\mathrm{D}_\ell$ can be obtained.

Using these two indicators, the primary locations for noise monitoring are pinpointed:
\begin{equation}
\ell^\star=\min\{\ell~|~\mathrm{M}_\ell>\overline{\mathrm{M}} \land\ \mathrm{D}_\ell>\overline{\mathrm{D}}\},
\end{equation}
$\ell^\star$ typically occurs in the later layers of the model.
Let $\mathcal{L}^\star = \{\ell^\star, \ell^\star+1, \dots, L\}$ denote the set of layers at which the basis is retained and used for identification.

\noindent \textbf{Noise Basis Extraction.}
For each $\ell \in \mathcal{L}^\star$, the semantic matrix $\mathbf{S}_\ell$ and noise matrix $\mathbf{N}_\ell$ are decomposed via singular value decomposition (SVD)~\cite{stewart1993early} as
$\mathbf{S}_\ell = \mathbf{U}^s_\ell \mathbf{\Sigma}^s_\ell (\mathbf{V}^s_\ell)^\top$ and
$\mathbf{N}_\ell = \mathbf{U}^n_\ell \mathbf{\Sigma}^n_\ell (\mathbf{V}^n_\ell)^\top$, respectively.

Here $\mathbf{U}^s_\ell,\mathbf{U}^n_\ell\in\mathbb{R}^{M\times M}$ span the sample space,
$\mathbf{V}^s_\ell,\mathbf{V}^n_\ell\in\mathbb{R}^{d_\ell\times d_\ell}$ span the hidden space,
and $\mathbf{\Sigma}^s_\ell,\mathbf{\Sigma}^n_\ell\in\mathbb{R}^{M\times d_\ell}$ store singular values in descending order.
Importantly, because our goal is to identify directions in the hidden space, we focus on the right singular vectors in $V$.
Each column $\mathbf{v}^s_{\ell,j}\in\mathbb{R}^{d_\ell}$ (resp. $\mathbf{v}^n_{\ell,j}$) represents a principal direction along which semantic (resp. noise) independent activations.

Let $\sigma^s_{\ell,j}$ (resp. $\sigma^n_{\ell,j}$) denote the $j$-th singular value associated with $\mathbf{v}^s_{\ell,j}$ (resp. $\mathbf{v}^n_{\ell,j}$).
Larger singular values indicate directions that contain stronger similarity information at layer $\ell$. We retain dominant semantic $\mathcal{I}^s_\ell=\{j~|~\sigma^s_{\ell,j}>\alpha\}$ and noise information  $\mathcal{I}^n_\ell=\{j~|~\sigma^n_{\ell,j}>\alpha\}$ via threshold $\alpha$.
In subsequent steps, we only use the corresponding right singular vectors $\{\mathbf{v}^s_{\ell,j}\}_{j \in \mathcal{I}^s_\ell}$ and $\{\mathbf{v}^n_{\ell,j}\}_{j \in \mathcal{I}^n_\ell}$ to construct the noise-only subspace for SEE.

For every noise direction index $j \in \mathcal{I}^n_{\ell}$, we compute its cosine similarity with each semantic direction $k \in \mathcal{I}^s_{\ell}$, and aggregate them by the maximum absolute similarity:
\begin{equation}
\textbf{m}_{\ell,j} =\max_{k\in\mathcal{I}^s_{\ell}}(\mathrm{cos}(\mathbf{v}^n_{{\ell},j},\mathbf{v}^s_{{\ell},k})).
\end{equation}

A dominant noise direction is retained if it is nearly orthogonal to all dominant semantic directions:
$\mathcal{J}_\ell=\Big\{j\in\mathcal{I}^n_\ell~\big|~\textbf{m}_{\ell,j} <\delta\Big\}$.

Define a binary mask vector $s_\ell \in \{0,1\}^{d_t}$ with $s_\ell[j]=1$ if $j \in \mathcal{J}_\ell$, and a diagonal mask matrix $\mathbf{M}_\ell=\text{diag}(s_\ell)\in \mathbb{R}^{d_\ell\times d_\ell}$. Then we get noise basis matrix:
\begin{equation}
\mathbf{Q}_\ell=\mathbf{V}_\ell^n\mathbf{M}_\ell
\in\mathbb{R}^{d_\ell\times r_\ell}.
\end{equation}

\begin{figure*}[t]
    \centering \includegraphics[width=1\textwidth]{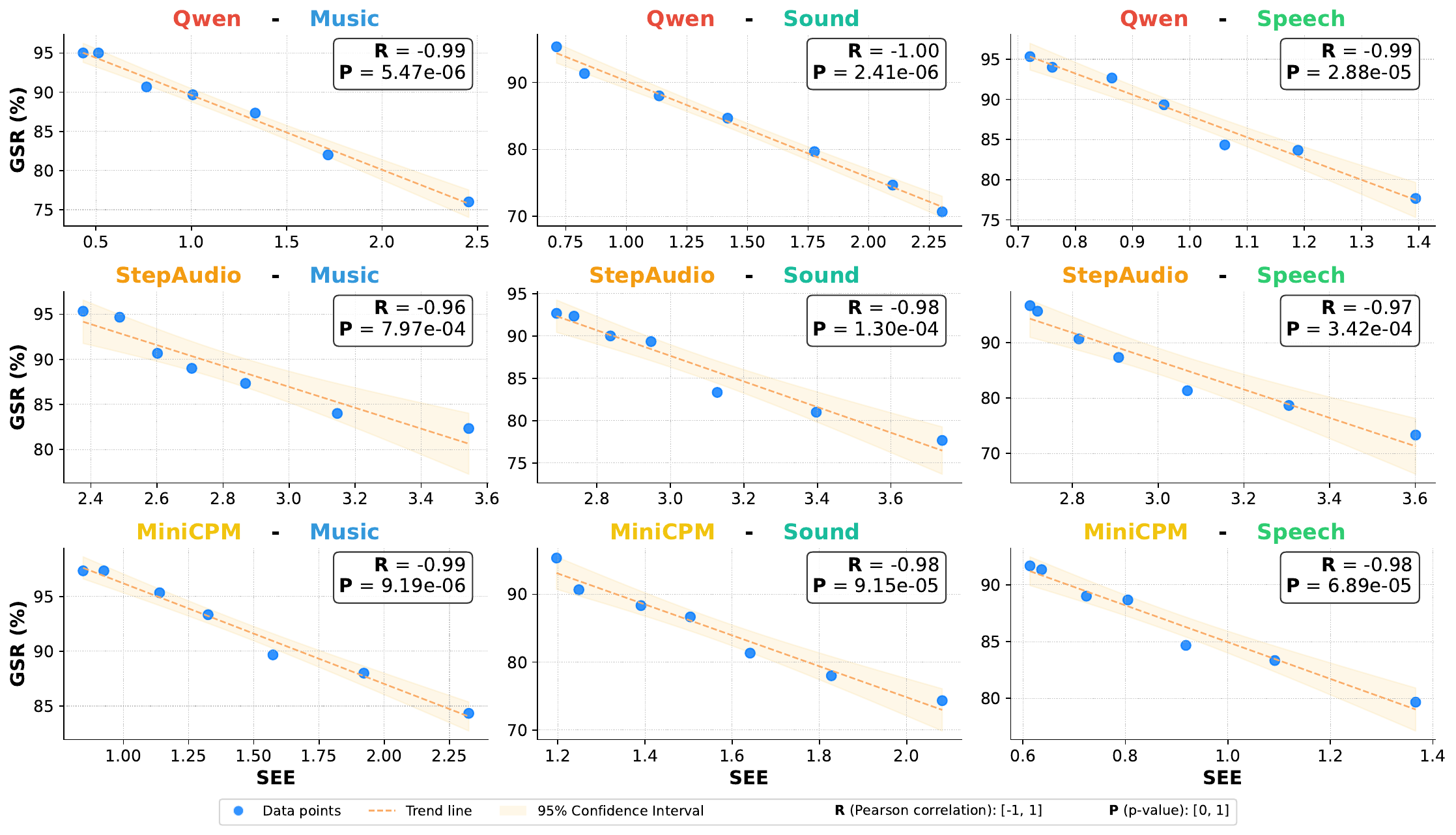}
    \caption{SEE is negatively correlated with GSR, with a consistent trend observed across different task types.}
    \label{fig:dependency_3_3}
\end{figure*}
\subsection{SEE Calculation}
For each audio input $x$, we project token activations onto the noise directions for each $\ell\in\mathcal{L}^\star$:
\begin{equation}
\mathbf{Z}_\ell(x)=\mathbf{A}_\ell(x)\mathbf{Q}_\ell \in\mathbb{R}^{T(x)\times r_\ell}.
\end{equation}

For each frame token $t$, we define the layer SEE component as the energy:
\begin{equation}
\text{SEE}_\ell(x)=\frac{1}{T(x)}\sum_{t=1}^{T(x)}\frac{\|\mathbf{Z}_\ell(x)_{t,:}\|_2^2}{\|\mathbf{A}_\ell(x)_{t,:}\|_2^2+\varepsilon}.
\end{equation}

Finally, we aggregate over retained layers to form a single score for the input:
\begin{equation}
\mathrm{SEE}(x)=\frac{1}{|\mathcal{L}^\star|}\sum_{\ell\in\mathcal{L}^\star} \text{SEE}_\ell(x),
\end{equation}
$\mathrm{SEE}(x)$ measures the embedding energy aligned with the noise subspace, and thus directly quantifies noise interference inside the model.

\subsection{Signal Embedding Energy Neutralization (SEEN) Construction}
While SEE quantifies how strongly an input activates noise directions, SEEN uses the same noise subspace to remove the corresponding components from intermediate activations, aiming to reduce noise bias without changing model parameters or requiring additional training data.

Given the retained layer set $\mathcal{L}^\star$ and the noise matrices $\{\mathbf{Q}_\ell\}_{\ell\in\mathcal{L}^\star}$, we manipulate the activation in the audio embedding. For an input $x$ and layer $\ell\in\mathcal{L}^\star$, we first reconstruct the noise component in the original hidden space by projecting $\mathbf{A}_\ell(x)\in\mathbb{R}^{T(x)\times d_\ell}$ onto the noise subspace:
\begin{equation}
\mathbf{C}_\ell(x)=\mathbf{A}_\ell(x)\mathbf{Q}_\ell\mathbf{Q}_\ell^\top \in \mathbb{R}^{T(x)\times d_\ell}.
\end{equation}

SEEN then neutralizes the activation by subtracting the projected component:
\begin{equation}
\begin{aligned}
&\widetilde{\mathbf{A}}_\ell(x)=\text{SEEN}(x)=\mathbf{A}_\ell(x)-\lambda \mathbf{C}_\ell(x),\\
&\lambda\in[0,1],\ \ell\in\mathcal{L}^\star.
\end{aligned}
\end{equation}
Here $\lambda$ controls the neutralization strength (the default setting is $\lambda=1$).

The wthen continues the forward pass using $\widetilde{\mathbf{A}}_\ell(x)$. In this way, SEEN provides an effective and lightweight means to align noise mitigation with SEE, thereby reducing internal perturbation while preserving the remaining semantic components needed for generation.

\section{Experiments}
\begin{table*}[t]
\centering
\resizebox{\textwidth}{!}{%
\begin{tabular}{@{}l|cccc|ccc@{}}
\toprule
\textbf{Noise}  \quad & \quad  {\ul \textbf{Gauss}}  \quad & \quad  {\ul \textbf{Crowd}}  \quad & \quad  {\ul \textbf{Traffic}}  \quad & \quad  {\ul \textbf{Machine}}  \quad & \quad  \textbf{Animal}  \quad & \quad  \textbf{Shower} \quad & \quad  \textbf{Wind}  \quad  \\ \midrule
\textbf{Clean} & 1.212 & 1.212 & 1.212 & 1.212 & 1.212 & 1.212 & 1.212 \\
\rowcolor[HTML]{EFEFEF} 
\textbf{10db} & 1.767 & 1.709 & 1.633 & 2.106 & 1.462 & 1.503 & 1.402 \\
\textbf{0db} & 2.052 & 2.287 & 2.324 & 2.808 & 1.544 & 1.536 & 1.551 \\
\rowcolor[HTML]{EFEFEF} 
\textbf{-10db} & 2.436 & 3.101 & 3.481 & 3.306 & 1.797 & 1.586 & 1.921 \\ \bottomrule
\end{tabular}%
}
\caption{The effect of SEE generated with mixed noise types. The underlined items are known noise types.}
\label{tab:cross_noise1}
\vspace{-5pt}
\end{table*}
\subsection{Setups}
\noindent \textbf{Models.}
We conduct experiments across 3 models, including Qwen (\texttt{Qwen-2.5-omni-7b})~\cite{Qwen2-Audio}, MiniCPM (\texttt{Minicpm-o-2.6})~\cite{yao2024minicpm}, and StepAudio (\texttt{Step-Audio-2-mini})~\cite{wu2025stepaudio2technicalreport}.
Additional experimental details are provided in the Appendix~\ref{app:more experiment}

\noindent \textbf{Datasets.}
In the experiments, we primarily evaluate our method on the MMAU~\cite{kumar2025mmau} and Librispeech~\cite{panayotov2015librispeech} datasets. 
Based on the underlying audio modality, the tasks are grouped into four categories: speech-to-text (STT), Speech, Sound, and Music. 
STT evaluates word-level recognition. 
Speech focuses on world-knowledge QA. 
Sound assesses environmental sound perception, while music examines multicultural music reasoning.

For noise settings, we employ randomly generated white Gaussian noise (Gauss)~\cite{ko2015audio}, together with noise categories from the PNL dataset~\cite{hu2010tandem}, including crowd noise (Crowd), mechanical noise (Machine), and vehicle noise (Traffic). 

\noindent \textbf{Baselines.} 
We measure the deviation between model outputs under normal (Clean) and noisy (Noise) conditions as our metric, rather than ground-truth accuracy.

For traditional audio denoising, we select two frequency-based and two model-based methods as baselines. 
Specifically, STFT~\cite{ephraim2003speech} and WT~\cite{donoho2002noising} represent frequency-based approaches, while Segan~\cite{pascual2017segan} and DFL~\cite{purwins2019deep} are adopted as model-based baselines.

\noindent \textbf{Metrics.} 
Noise intensity is standardized using SNR, set to $-10$ dB for unlabeled experiments. We use Generation Success Rate (GSR) as the primary metric, measuring noisy sample accuracy relative to clean inputs. The SEE index's validity is assessed via Pearson correlation~\cite{benesty2009pearson} with GSR and associated p-values. 

\subsection{Quantifying Noise Intensity with SEE}
\begin{table}[t]
\centering
\resizebox{\columnwidth}{!}{%
\begin{tabular}{@{}l|l|cccc@{}}
\toprule
\textbf{Dataset} & \textbf{Method} & \textbf{Gauss} & \textbf{Crowd} & \textbf{Machine} & \textbf{Traffic} \\ \midrule
 & \textbf{Clean} & 0.35 & 1.32 & 1.30 & 1.17 \\
\multirow{-2}{*}{\textbf{Music}} & \cellcolor[HTML]{EFEFEF}\textbf{Noise} & \cellcolor[HTML]{EFEFEF}2.45 & \cellcolor[HTML]{EFEFEF}5.97 & \cellcolor[HTML]{EFEFEF}11.15 & \cellcolor[HTML]{EFEFEF}5.84 \\
 & \textbf{Clean} & 0.67 & 2.12 & 2.47 & 1.85 \\
\multirow{-2}{*}{\textbf{Sound}} & \cellcolor[HTML]{EFEFEF}\textbf{Noise} & \cellcolor[HTML]{EFEFEF}2.72 & \cellcolor[HTML]{EFEFEF}5.71 & \cellcolor[HTML]{EFEFEF}10.87 & \cellcolor[HTML]{EFEFEF}5.30 \\
 & \textbf{Clean} & 0.66 & 3.12 & 3.54 & 3.28 \\
\multirow{-2}{*}{\textbf{Speech}} & \cellcolor[HTML]{EFEFEF}\textbf{Noise} & \cellcolor[HTML]{EFEFEF}1.63 & \cellcolor[HTML]{EFEFEF}4.79 & \cellcolor[HTML]{EFEFEF}6.74 & \cellcolor[HTML]{EFEFEF}4.14 \\ \bottomrule
\end{tabular}%
}
\caption{SEE separation across noise types. }
\label{tab:more_noice}
\end{table}
\begin{figure}[t]
    \centering \includegraphics[width=1\columnwidth]{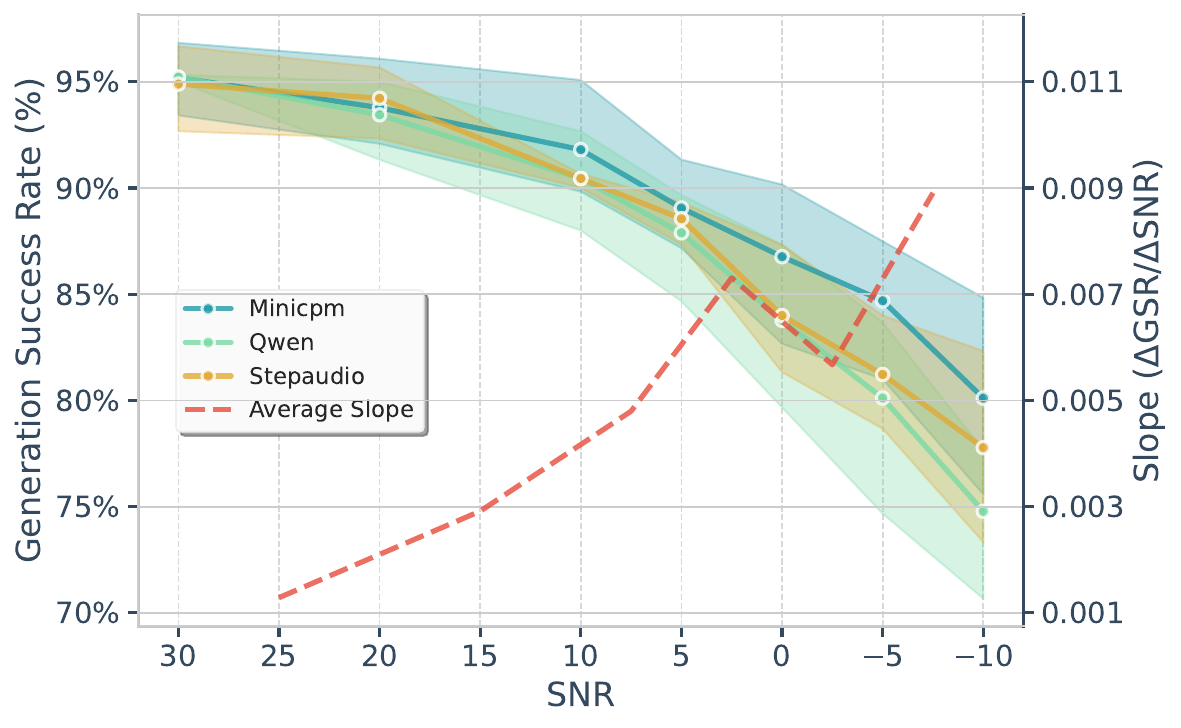}
    \caption{Generation success rate (GSR) versus SNR, showing a sharp degradation below $\text{SNR}=10$.}
    \label{fig:ASR_on_SNR_line}
\end{figure}
We evaluate Signal Embedding Energy (SEE) across three LALM frameworks to assess their robustness to noise.
As shown in Figure~\ref{fig:dependency_3_3}, GSR degrades consistently as SEE increases, with strong Pearson correlations ($R \in [-0.96, -1.00], P \ll 0.001$), indicating that SEE is tightly coupled with performance drop under corruption. This confirms SEE as a model-centric metric that captures embedding-space interference to predict generation failure. Additional empirical evidence is provided in the Appendix ~\ref{app:Supplementary Analysis of SEE} 

The above analysis establishes SEE as an internal indicator; we also evaluate its robustness against a standard physical SNR scale to validate the selected GSR interval. Figure~\ref{fig:ASR_on_SNR_line} shows robustness versus SNR. Across all LALMs, GSR degrades monotonically as SNR decreases, falling more steeply below $\approx 10$dB, which marks an SNR sensitivity regime. The slope curve confirms this transition via larger marginal losses at low SNR. This physical turning point aligns with the GSR range where SEE rises in our previous analysis, confirming that our selected interval effectively captures noise-induced failures.

\subsection{Effectiveness of SEE Across Noise Types}
\begin{table*}[t]
\centering
\resizebox{\textwidth}{!}{%
\begin{tabular}{@{}l|l|cc|cc|cc@{}}
\toprule
 &  & \multicolumn{2}{c|}{\textbf{MiniCPM}} & \multicolumn{2}{c|}{\textbf{Qwen}} & \multicolumn{2}{c}{\textbf{StepAudio}} \\ \cmidrule(l){3-8} 
\multirow{-2}{*}{\textbf{Dataset}} & \multirow{-2}{*}{\textbf{Method}} & \quad  \quad  \textbf{Noise} \quad  \quad  & \quad  \quad  \textbf{Clean} \quad  \quad   & \quad  \quad  \textbf{Noise} \quad  \quad  & \quad  \quad  \textbf{Clean} \quad  \quad  & \quad  \quad  \textbf{Noise} \quad  \quad  & \quad  \quad  \textbf{Clean} \quad  \quad  \\ \midrule
 & \textbf{STFT} & 84.33\% & 93.33\% & 71.67\% & 90.33\% & 79.00\% & 92.67\% \\
 & \cellcolor[HTML]{EFEFEF}\textbf{WT} & \cellcolor[HTML]{EFEFEF}82.00\% & \cellcolor[HTML]{EFEFEF}94.67\% & \cellcolor[HTML]{EFEFEF}67.67\% & \cellcolor[HTML]{EFEFEF}91.00\% & \cellcolor[HTML]{EFEFEF}77.67\% & \cellcolor[HTML]{EFEFEF}94.00\% \\
 & \textbf{Segan} & 82.67\% & 89.67\% & 71.00\% & 86.33\% & 78.00\% & 87.33\% \\
 & \cellcolor[HTML]{EFEFEF}\textbf{DFL} & \cellcolor[HTML]{EFEFEF}82.00\% & \cellcolor[HTML]{EFEFEF}88.33\% & \cellcolor[HTML]{EFEFEF}67.33\% & \cellcolor[HTML]{EFEFEF}76.67\% & \cellcolor[HTML]{EFEFEF}75.67\% & \cellcolor[HTML]{EFEFEF}85.33\% \\
\multirow{-5}{*}{\textbf{Music}} & \textbf{SEEN} & \textbf{85.00\%} & \textbf{99.00\%} & \textbf{76.99\%} & \textbf{97.67\%} & \textbf{82.67\%} & \textbf{98.00\%} \\ \midrule
 & \textbf{STFT} & 72.00\% & 86.67\% & 64.33\% & 88.33\% & 76.00\% & 87.67\% \\
 & \cellcolor[HTML]{EFEFEF}\textbf{WT} & \cellcolor[HTML]{EFEFEF}70.33\% & \cellcolor[HTML]{EFEFEF}92.33\% & \cellcolor[HTML]{EFEFEF}63.00\% & \cellcolor[HTML]{EFEFEF}91.67\% & \cellcolor[HTML]{EFEFEF}71.33\% & \cellcolor[HTML]{EFEFEF}92.33\% \\
 & \textbf{Segan} & 73.00\% & 87.00\% & 58.67\% & 83.33\% & 73.67\% & 87.00\% \\
 & \cellcolor[HTML]{EFEFEF}\textbf{DFL} & \cellcolor[HTML]{EFEFEF}73.67\% & \cellcolor[HTML]{EFEFEF}81.33\% & \cellcolor[HTML]{EFEFEF}60.00\% & \cellcolor[HTML]{EFEFEF}79.33\% & \cellcolor[HTML]{EFEFEF}72.33\% & \cellcolor[HTML]{EFEFEF}83.00\% \\
\multirow{-5}{*}{\textbf{Sound}} & \textbf{SEEN} & \textbf{75.27\%} & \textbf{98.67\%} & \textbf{72.00\%} & \textbf{99.67\%} & \textbf{78.33\%} & \textbf{99.00\%} \\ \midrule
 & \textbf{STFT} & 75.33\% & 94.00\% & 68.00\% & 97.00\% & 66.33\% & 93.33\% \\
 & \cellcolor[HTML]{EFEFEF}\textbf{WT} & \cellcolor[HTML]{EFEFEF}70.67\% & \cellcolor[HTML]{EFEFEF}93.00\% & \cellcolor[HTML]{EFEFEF}68.00\% & \cellcolor[HTML]{EFEFEF}97.00\% & \cellcolor[HTML]{EFEFEF}55.67\% & \cellcolor[HTML]{EFEFEF}94.67\% \\
 & \textbf{Segan} & 69.00\% & 93.00\% & 64.67\% & 91.33\% & 60.67\% & 90.00\% \\
 & \cellcolor[HTML]{EFEFEF}\textbf{DFL} & \cellcolor[HTML]{EFEFEF}72.67\% & \cellcolor[HTML]{EFEFEF}93.00\% & \cellcolor[HTML]{EFEFEF}65.33\% & \cellcolor[HTML]{EFEFEF}92.00\% & \cellcolor[HTML]{EFEFEF}62.00\% & \cellcolor[HTML]{EFEFEF}92.33\% \\
\multirow{-5}{*}{\textbf{Speech}} & \textbf{SEEN} & \textbf{80.00\%} & \textbf{98.00\%} & \textbf{78.33\%} & \textbf{99.00\%} & \textbf{74.00\%} & \textbf{96.33\%} \\ \bottomrule
\end{tabular}%
}
\caption{Generation success rate under noisy conditions with different denoising strategies.}
\label{tab:SEEN_ASR}
\end{table*}
We evaluated SEE across noise types to verify generalizability. Results on Qwen (Table~\ref{tab:more_noice}) show a clear gap between clean and noisy conditions. Specifically, clean and noisy inputs exhibit non-overlapping SEE ranges across request types, with SEE increasing for all noise categories.

Furthermore, we evaluate the applicability of SEE across diverse noise categories and optimize SEE under four representative noise types (Gauss, Crowd, Machine, and Traffic).
Table~\ref{tab:cross_noise1} shows consistent SEE values across these categories at the same SNR, suggesting SEE is driven by noise intensity rather than specific acoustic traits.

To further examine the generalization capability, we introduce three additional noise categories that are not included in the optimization process. Table~\ref{tab:cross_noise1} reported slightly lower SEE values compared to those obtained on directly optimized noise types, likely due to differing spectral and temporal structures. However, SEE still increases monotonically as SNR drops. This preserved monotonicity confirms that SEE remains sensitive to noise intensity even under unseen conditions. More detailed results are reported in Appendix~\ref{Sec:cross-over}

\subsection{SEE on traditional denoising methods}
\begin{figure}[t]
    \centering \includegraphics[width=1\columnwidth]{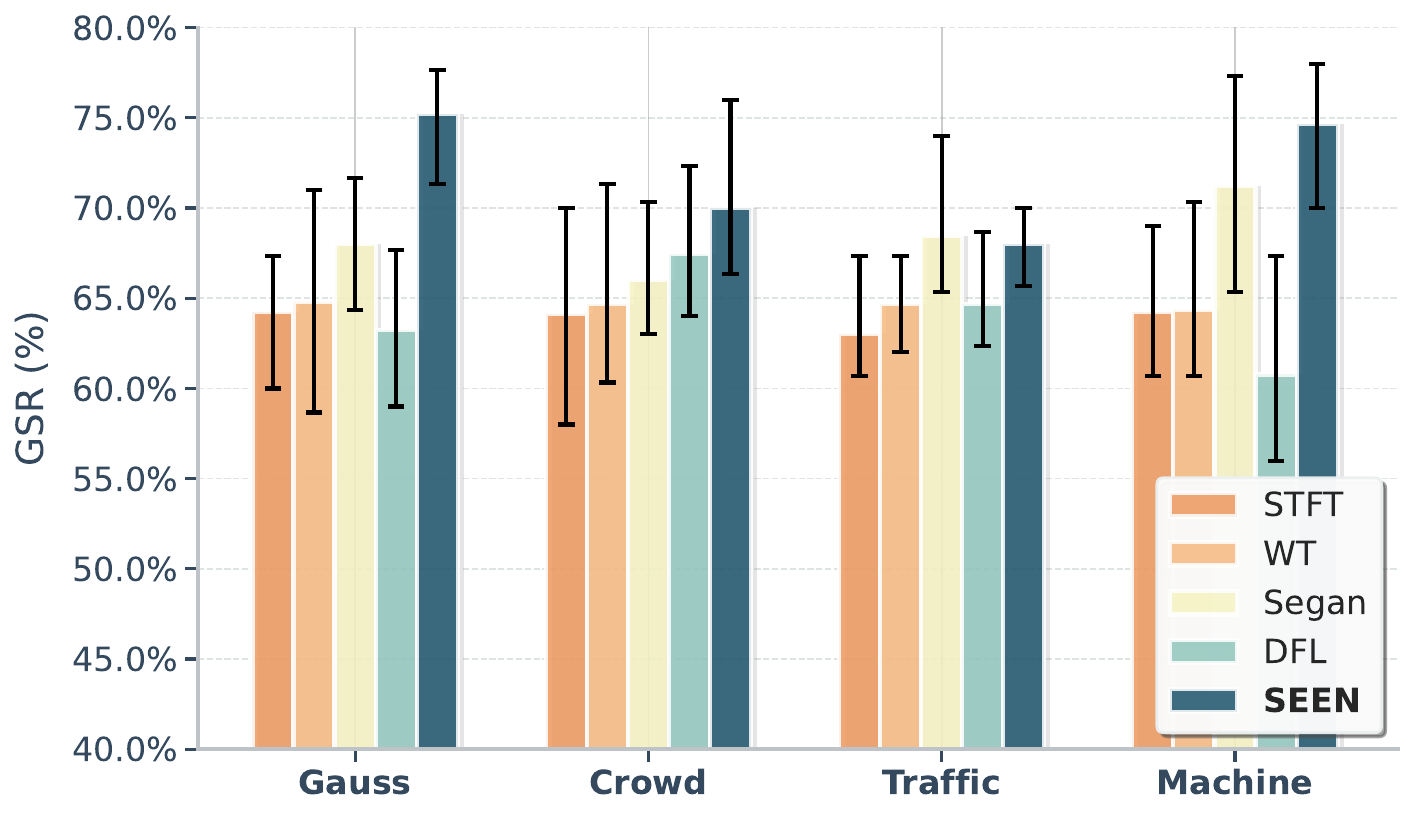}
    \caption{SEEN improves generation quality across noise types without consistent performance degradation.}
    \label{fig:More_noise_ASR}
\end{figure}
\begin{figure}[t]
    \centering \includegraphics[width=1\columnwidth]{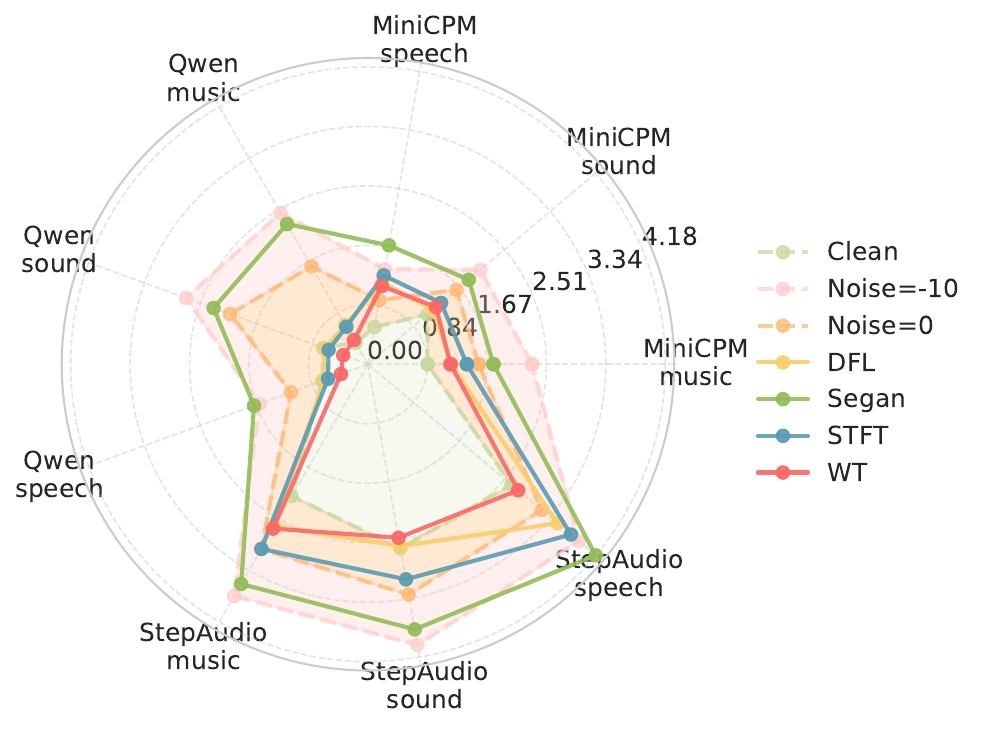}
    \caption{Effect of waveform denoising measured by SEE. Many conventional denoisers fail to reduce SEE and can even increase it. Lower SEE indicates less noise-aligned interference.
}
    \label{fig:SEE_radar_chart}
    \vspace{-5pt}
\end{figure}

We use SEE to analyze preprocessed samples. As shown in Figure~\ref{fig:SEE_radar_chart}, denoised inputs generally exhibit higher SEE than raw inputs. This contrasts with waveform-level evidence that these methods reduce acoustic noise, highlighting a mismatch between denoising objectives and LALM semantic encoding. Quantitatively, the SEE of denoised audio is comparable to inputs with an additional $\text{SNR}=0$, suggesting that denoising artifacts act like extra noise within the model's internal representations. 

Moreover, denoising causes unique waveform misalignments with clean embeddings. While SEE captures residual interference, these distortions extend beyond the original noise. This indicates added variability in the encoding space that may hinder robustness-oriented training. More results are reported in Appendices~\ref{app:Evaluate SEEN by SEE} and ~\ref{app:sim}

\subsection{Suppressing Noise with SEEN}
\begin{table}[t]
\centering
\resizebox{\columnwidth}{!}{%
\begin{tabular}{@{}l|l|cccc@{}}
\toprule
\textbf{Model} & \textbf{Dataset} & \textbf{Early} & \textbf{Middle} & \textbf{ALL} & \textbf{SEEN} \\ \midrule
\textbf{MiniCPM} & \textbf{Clean} & 0.002 & 0.004 & 0.004 & 0.763 \\
\textbf{} & \cellcolor[HTML]{EFEFEF}\textbf{Noise} & \cellcolor[HTML]{EFEFEF}0.126 & \cellcolor[HTML]{EFEFEF}0.270 & \cellcolor[HTML]{EFEFEF}0.456 & \cellcolor[HTML]{EFEFEF}1.777 \\ \midrule
\textbf{Qwen} & \textbf{Clean} & 0.001 & 0.000 & 0.015 & 0.633 \\
\textbf{} & \cellcolor[HTML]{EFEFEF}\textbf{Noise} & \cellcolor[HTML]{EFEFEF}0.024 & \cellcolor[HTML]{EFEFEF}0.005 & \cellcolor[HTML]{EFEFEF}1.751 & \cellcolor[HTML]{EFEFEF}1.972 \\ \midrule
\textbf{StepAudio} & \textbf{Clean} & 0.005 & 0.022 & 0.092 & 2.489 \\
\textbf{} & \cellcolor[HTML]{EFEFEF}\textbf{Noise} & \cellcolor[HTML]{EFEFEF}0.312 & \cellcolor[HTML]{EFEFEF}0.661 & \cellcolor[HTML]{EFEFEF}5.233 & \cellcolor[HTML]{EFEFEF}3.878 \\ \bottomrule
\end{tabular}%
}
\caption{Ablation on layer selection for SEE. }
\label{tab:Ablation_1}
\end{table}
\begin{table}[t]
\centering
\resizebox{\columnwidth}{!}{%
\begin{tabular}{@{}l|ccc@{}}
\toprule
\textbf{Model}  \quad & \quad  \textbf{MiniCPM} \quad  &  \quad \textbf{Qwen} \quad  &  \quad \textbf{StepAudio}  \quad \\ \midrule
\textbf{None} & 84.33\% & 76.00\% & 82.33\% \\
\rowcolor[HTML]{EFEFEF} 
\textbf{0.25} & 84.67\% & 76.33\% & 82.67\% \\
\textbf{0.5} & 84.33\% & 76.33\% & 82.67\% \\
\rowcolor[HTML]{EFEFEF} 
\textbf{0.75} & 85.00\% & 76.67\% & 82.67\% \\
\textbf{1} & \textbf{85.00\%} & \textbf{77.00\%} & \textbf{82.67\%} \\
\rowcolor[HTML]{EFEFEF} 
\textbf{1.2} & 85.00\% & 76.67\% & 82.67\% \\ \bottomrule
\end{tabular}%
}
\caption{Sensitivity of SEEN to the neutralization strength $\lambda$. }
\label{tab:Ablation_a}
\end{table}
\begin{table}[t]
\centering
\resizebox{\columnwidth}{!}{%
\begin{tabular}{@{}l|ccc@{}}
\toprule
\textbf{Model} \quad  & \quad \textbf{MiniCPM}  \quad & \quad  \textbf{Qwen}  \quad & \quad  \textbf{StepAudio} \quad  \\ \midrule
\textbf{None} & 84.33\% & 76.00\% & 82.33\% \\
\rowcolor[HTML]{EFEFEF} 
\textbf{0.1} & \textbf{85.00\%} & \textbf{77.00\%} & \textbf{82.67\%} \\
\textbf{0.3} & 84.33\% & 77.00\% & 82.33\% \\
\rowcolor[HTML]{EFEFEF} 
\textbf{0.5} & 83.33\% & 77.33\% & 82.33\% \\
\textbf{0.7} & 84.33\% & 77.33\% & 82.33\% \\
\rowcolor[HTML]{EFEFEF} 
\textbf{1} & 83.67\% & 77.33\% & 82.33\% \\ \bottomrule
\end{tabular}%
}
\caption{Sensitivity of SEEN to the cosine threshold $\delta$.}
\label{tab:Ablation_lamada}
\end{table}
Table~\ref{tab:SEEN_ASR} evaluates SEEN's impact on generation quality. In clean or low-noise settings, SEEN causes negligible performance drops and preserves semantic fidelity, confirming that suppressing noise-related directions does not harm normal requests. Under noise, SEEN consistently improves GSR across all datasets and models, outperforming traditional and learned denoisers by 6.7\% on average. These results suggest that embedding-space mitigation is more effective than acoustic suppression, as it avoids new distortions while targeting noise-induced semantic bias. Appendix~\ref{Sec:SEEN_validity_statement} provides more analysis.

We also assess SEEN's robustness across noise types on three MMAU datasets. As shown in Table~\ref{fig:More_noise_ASR}, SEEN maintains effective noise suppression across all tested conditions.
Additional results of the SEEN method under various SNR settings are reported in Appendix~\ref{app:Summary of the Performance of SEEN under Various SNRs}.
\subsection{Ablation Analysis}
\noindent \textbf{Layer Selection for Noise Monitoring.}
We first ablate the layer-selection strategy used in SEE. 
To evaluate its necessity, we construct three alternatives: (i) applying the method to the first third of layers (Early),
(ii) only to the middle third of layers (Middle), and
(iii) to all layers(ALL).

Table ~\ref{tab:Ablation_1} shows that restricting SEE to the early layers or middle layers does not produce significant results.
This is because screening at these layers produces few valid directions, leading to an unstable detection.
Applying the method to all layers is feasible. However, in different models, the variation magnitude becomes unstable and can be influenced by interference from early and middle layers.
Our method is more closely aligned with the extracted noise subspace and better preserves cross-modal embedding consistency.

\noindent \textbf{Sensitivity to Hyperparameters.}
We next study the sensitivity of SEEN to two key hyperparameters: the suppression strength $\lambda$ and the cosine similarity threshold $\delta$.
Table~\ref{tab:Ablation_a} shows that larger values of $\lambda$ lead to more effective suppression, and that strengthening noise-direction suppression improves generation quality. However, when the suppression ratio exceeds 1, components opposite to the noise direction are introduced, causing additional semantic perturbations.
Table~\ref{tab:Ablation_lamada} reports the effect of cosine similarity. In most cases, amplifying the $\delta$ interferes with the main semantic direction, ultimately degrading performance.

\section{Conclusion}
We introduce Signal Embedding Energy (SEE), a metric to quantify real-world noise impact. 
SEE is based on structured activation subspaces derived from the model’s internal representations, enabling a more accurate characterization of noise perception than clean audio features.
Using SEE as a probe, we show that state-of-the-art LALMs are highly sensitive to noise. More critically, conventional speech denoising techniques often fail to reduce SEE, revealing a semantic misalignment between acoustic suppression and LALM embedding-level reasoning. Driven by SEE, we propose Signal Embedding Energy Neutralization (SEEN), a training-free method that removes noise subspaces from embedding activations. SEEN consistently reduces SEE and outperforms existing denoising methods. Overall, our study highlights the need for semantic-level criteria and offering a principled direction for future LALM design.

\section*{Limitations}
Our approach assumes access to aligned clean requests and pure-noise recordings from the target deployment environment to estimate a stable noise subspace. In practice, such "noise-only" collections cannot be directly acquired during training; they require collecting noise from the actual application environment. 
Downstream application environments with relatively single or few categories of noise are easier to collect and are suitable for SEE measurement.
Highly variable environments can increase the difficulty of obtaining pure-noise recordings.

Methodologically, our current SEE compresses variable length via mean pooling, which may underestimate temporally corrupted and paralinguistic cues. While effective in our tested regimes, overly aggressive neutralization could remove task-relevant information under certain acoustic conditions. A better approach would not be to eliminate noise semantic components, but rather to enhance the model's understanding of incomplete task semantics.

A promising direction is to use SEE as a training robustness signal rather than only a mitigation tool. 
This can be achieved by incorporating SEE as a regularizer under noise augmentation to explicitly penalize noise-aligned energy in embeddings, or by using SEE to select or learn embedding-space enhancement modules that reduce semantic interference without relying solely on acoustic fidelity. We leave this direction for future work.


\bibliography{custom}
\newpage
\appendix

\section{More Experimental Setups}
\label{app:more experiment}
In this section, we provide a comprehensive description of the experimental configurations used in our study, covering both the classification and transcription tasks.

\subsection{Model Architecture and Audio Processing}
All experiments utilize an omni-modal Large Language Model (LLM) loaded in \texttt{bfloat16} precision. We leverage the \texttt{sdpa} (Scaled Dot Product Attention) implementation for efficient computation. For acoustic input, all raw audio signals are resampled to 16,000 Hz and converted to mono-channel using the \texttt{librosa} library. All experiments are conducted on a server equipped with an NVIDIA RTX 5090 GPU.

\subsection{Task Definitions and Prompting}
We evaluate our method on two distinct types of tasks:
\begin{enumerate}
    \item \textbf{Multiple-Choice Question Answering (MCQA)}: Applied to \textit{Music}, \textit{Sound}, and \textit{Speech} datasets. The model is prompted to answer multiple-choice questions with a strict constraint: \textit{``Answer the multiple-choice question by outputting ONLY one letter.''} A regex pattern \texttt{r'([A-Za-z])'} is used to parse the final answer.
    \item \textbf{Speech-to-Text(STT)}: Applied to the \textit{LibriSpeech} dataset. The model is tasked with transcription using the prompt: \textit{``Please listen to the audio snippet carefully and transcribe the content.''}
\end{enumerate}

\subsection{Noise Synthesis and Dataset Scale}
To simulate real-world interference, we inject various types of environmental noise into the clean audio samples at seven SNR levels: $\{-10, -5, 0, 5, 10, 20, 30\}$. The sample sizes are set to $N=300$ for MCQA task and $N=100$ for STT task to ensure statistically significant results.
 
\subsection{SEE Intervention and Hyperparameters}
The SEE is implemented by registering forward hooks on the target layers. The set of target layers for intervention is identified through the diagnostic procedure outlined in \autoref{alg:see_setup}. 
\begin{itemize}
    \item \textbf{Target Layers}: The suppression is applied to the encoder layers of models. For Qwen, target layer is from layer 23 to the final layer (\texttt{model.thinker.audio_tower.layers[23:]}). For MiniCPM, target layer is from layer 18 to the final layer (\texttt{model.apm.layers[18:]}). For StepAudio, target layer is from layer 27 to the final layer (\texttt{model.llm.encoder.blocks[27:]})
    \item \textbf{Suppression Factor}: The intervention strength $\alpha=1.0$. Cosine similarity threshold $\delta=0.1$. The energy threshold ratio used to determine the number of retained singular values $\rho=0.95$.
    \item \textbf{Noise Subspace ($V_{noise}$)}: The noise singular vectors are derived during an offline calibration phase using a representative set of 50 pure-noise segments sampled from the speech dataset, with the intensity standardized to a reference level equivalent to 0 dB SNR.
\end{itemize}
\subsection{About Baseline Methods}
To ensure a fair and rigorous comparison, all traditional baseline methods included in this study were implemented and configured in strict accordance with the parameter settings and experimental protocols specified in their original publications. No additional hyperparameter tuning was performed on these baselines beyond the recommendations provided by the respective authors to maintain the integrity of the comparative analysis.

\section{Supplementary Analysis of SEE}
\label{app:Supplementary Analysis of SEE}
\begin{figure*}[t]
    \centering \includegraphics[width=1\textwidth]{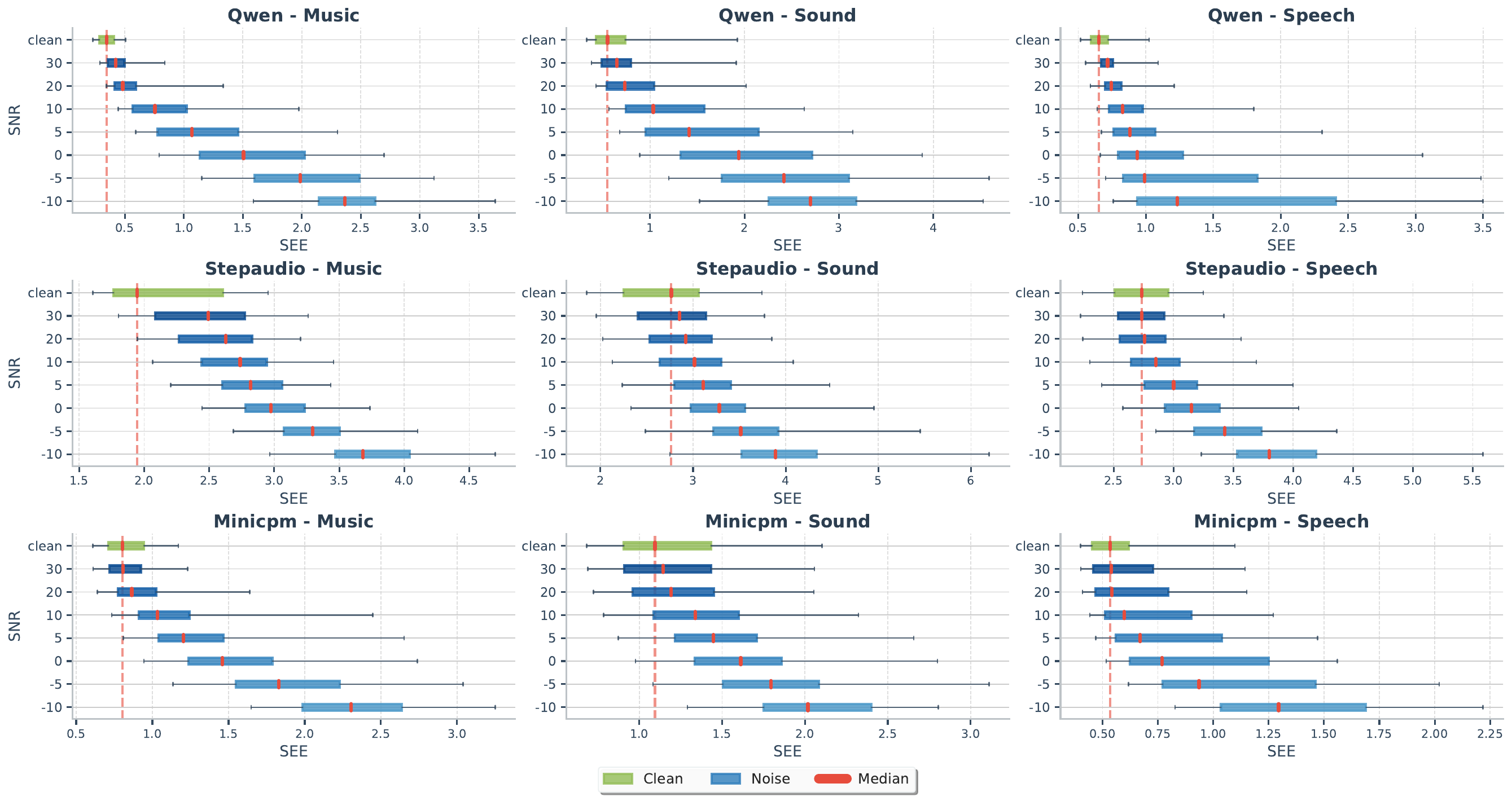}
    \caption{SEE under varying noise intensity across three LALMs. Clean inputs yield stable SEE, while SEE increases monotonically as SNR decreases. Boxplots report the interquartile range and median.}
    \label{fig:1_SEE_use}
\end{figure*}
\begin{table*}[t]
\centering
\resizebox{\textwidth}{!}{%
\begin{tabular}{@{}l|cccc|cccc|cccc@{}}
\toprule
 & \multicolumn{4}{c|}{\textbf{MiniCPM}} & \multicolumn{4}{c|}{\textbf{Qwen}} & \multicolumn{4}{c}{\textbf{StepAudio}} \\ \cmidrule(l){2-13} 
\multirow{-2}{*}{\textbf{Method}} & \textbf{STT} & \textbf{Music} & \textbf{Sound} & \textbf{Speech} & \textbf{STT} & \textbf{Music} & \textbf{Sound} & \textbf{Speech} & \textbf{STT} & \textbf{Music} & \textbf{Sound} & \textbf{Speech} \\ \midrule
\textbf{BASE} & 0.481 & 0.834 & 1.176 & 0.561 & 0.854 & 0.354 & 0.669 & 0.657 & 2.409 & 2.137 & 2.676 & 2.735 \\
\rowcolor[HTML]{EFEFEF} 
\textbf{Noise} & 1.378 & 2.306 & 2.064 & 1.357 & 1.082 & 2.455 & 2.725 & 1.625 & 3.872 & 3.763 & 4.007 & 3.871 \\
\textbf{STFT} & {\ul 1.416} & {\ul 1.399} & {\ul 1.347} & {\ul 1.318} & 0.607 & {\ul 0.604} & 0.599 & 0.600 & {\ul 4.442} & {\ul 3.049} & {\ul 3.025} & {\ul 3.658} \\
\rowcolor[HTML]{EFEFEF} 
\textbf{WT} & {\ul 1.188} & {\ul 1.213} & {\ul 1.255} & {\ul 1.153} & 0.442 & {\ul 0.391} & 0.369 & 0.404 & {\ul 3.001} & {\ul 2.667} & 2.476 & {\ul 2.756} \\
\textbf{Segan} & {\ul 1.906} & {\ul 1.816} & {\ul 1.910} & {\ul 1.682} & {\ul 2.051} & {\ul 2.330} & {\ul 2.270} & {\ul 1.675} & {\ul 5.078} & {\ul 3.829} & {\ul 3.698} & {\ul 4.014} \\
\rowcolor[HTML]{EFEFEF} 
\textbf{DFL} & {\ul 1.237} & {\ul 1.194} & {\ul 1.206} & {\ul 1.171} & 0.604 & {\ul 0.614} & 0.647 & {\ul 0.683} & {\ul 3.628} & {\ul 2.703} & {\ul 2.714} & {\ul 3.406} \\
\textbf{SEEN} & \textbf{0.008} & \textbf{0.017} & \textbf{0.016} & \textbf{0.009} & \textbf{0.007} & \textbf{0.015} & \textbf{0.017} & \textbf{0.010} & \textbf{0.040} & \textbf{0.044} & \textbf{0.047} & \textbf{0.039} \\ \bottomrule
\end{tabular}%
}
\caption{SEE scores under different settings. SEEN consistently reduces SEE, whereas most waveform denoisers increase SEE to varying degrees (underlined).}
\label{tab:SEEN_SEE}
\end{table*}
\noindent \textbf{SEE Is Stable on Clean Audio.}
We evaluate Signal Embedding Energy (SEE) across three LALM frameworks. 
Figure~\ref{fig:1_SEE_use} shows that,for clean audio, SEE remains stable within each model and exhibits only minor variation across categories, suggesting that the metric is largely insensitive to content differences when the input is clean. The clean SEE embeddings are concentrated in a relatively compact region, with limited activation along the noise-aligned subspace.

\noindent \textbf{SEE Tracks Noise-induced Degradation.}
As the signal-to-noise ratio (SNR) decreases, the injected noise becomes stronger and SEE increases monotonically, reflecting progressively larger projection onto the noise directions. The corresponding impact on generation is summarized in Figure~\ref{fig:ASR_on_SNR_line}: once SNR falls below 10 dB, task performance drops by more than 10\%, and under severe corruption $\text{SNR}=-10$ accuracy degrades by up to 25\%. This degradation aligns closely with the behavior of SEE, which rises sharply in the same SNR regime and becomes clearly separated from the clean-audio baseline. Overall, these results indicate that SEE reliably tracks noise intensity and captures how increasing acoustic corruption translates into interference.

\section{Evaluate SEEN by SEE}
\label{app:Evaluate SEEN by SEE}
We apply SEEN as an embedding-space denoising module and report its effect on SEE in Table~\ref{tab:SEEN_SEE}. In contrast to waveform-level denoisers, SEEN yields a substantial decrease in SEE. 
Because SEEN and SEE share similar computational structures, SEEN primarily suppresses components that are directly captured by SEE, which limits the interpretability of direct evaluation.

\section{Summary of the Performance of SEEN under Various SNRs}
\label{app:Summary of the Performance of SEEN under Various SNRs}


\begin{table*}[t]
\centering
\resizebox{\textwidth}{!}{%
\begin{tabular}{@{}l|l|cc|cc|cc@{}}
\toprule
\multirow{2}{*}{\textbf{Dataset}} & \multirow{2}{*}{\textbf{Method}} & \multicolumn{2}{c|}{\textbf{MiniCPM}} & \multicolumn{2}{c|}{\textbf{Qwen}} & \multicolumn{2}{c}{\textbf{StepAudio}} \\ \cmidrule(l){3-8} 
 &  & \quad \quad \textbf{Noise} \quad \quad & \quad \quad \textbf{Clean} \quad \quad & \quad \quad \textbf{Noise} \quad \quad & \quad \quad \textbf{Clean} \quad \quad & \quad \quad \textbf{Noise} \quad \quad & \quad \quad \textbf{Clean} \quad \quad \\ \midrule
\multirow{6}{*}{\textbf{Music}} & \textbf{None} & 84.33\% & 100.00\% & 76.00\% & 100.00\% & 82.33\% & 100.00\% \\
 & \textbf{STFT} & 84.33\% & 93.33\% & 71.67\% & 90.33\% & 79.00\% & 92.67\% \\
 & \textbf{WT} & 82.00\% & 94.67\% & 67.67\% & 91.00\% & 77.67\% & 94.00\% \\
 & \textbf{Segan} & 82.67\% & 89.67\% & 71.00\% & 86.33\% & 78.00\% & 87.33\% \\
 & \textbf{DFL} & 82.00\% & 88.33\% & 67.33\% & 76.67\% & 75.67\% & 85.33\% \\
 & \textbf{SEEN} & \textbf{85.00\%} & \textbf{99.00\%} & \textbf{76.99\%} & \textbf{97.67\%} & \textbf{82.67\%} & \textbf{98.00\%} \\ \midrule
\multirow{6}{*}{\textbf{Sound}} & \textbf{None} & 74.33\% & 100.00\% & 70.67\% & 100.00\% & 77.67\% & 100.00\% \\
 & \textbf{STFT} & 72.00\% & 86.67\% & 64.33\% & 88.33\% & 76.00\% & 87.67\% \\
 & \textbf{WT} & 70.33\% & 92.33\% & 63.00\% & 91.67\% & 71.33\% & 92.33\% \\
 & \textbf{Segan} & 73.00\% & 87.00\% & 58.67\% & 83.33\% & 73.67\% & 87.00\% \\
 & \textbf{DFL} & 73.67\% & 81.33\% & 60.00\% & 79.33\% & 72.33\% & 83.00\% \\
 & \textbf{SEEN} & \textbf{75.27\%} & \textbf{98.67\%} & \textbf{72.00\%} & \textbf{99.67\%} & \textbf{78.33\%} & \textbf{99.00\%} \\ \midrule
\multirow{6}{*}{\textbf{Speech}} & \textbf{None} & 79.67\% & 100.00\% & 77.67\% & 100.00\% & 73.33\% & 100.00\% \\
 & \textbf{STFT} & 75.33\% & 94.00\% & 68.00\% & 97.00\% & 66.33\% & 93.33\% \\
 & \textbf{WT} & 70.67\% & 93.00\% & 68.00\% & 97.00\% & 55.67\% & 94.67\% \\
 & \textbf{Segan} & 69.00\% & 93.00\% & 64.67\% & 91.33\% & 60.67\% & 90.00\% \\
 & \textbf{DFL} & 72.67\% & 93.00\% & 65.33\% & 92.00\% & 62.00\% & 92.33\% \\
 & \textbf{SEEN} & \textbf{80.00\%} & \textbf{98.00\%} & \textbf{78.33\%} & \textbf{99.00\%} & \textbf{74.00\%} & \textbf{96.33\%} \\ \bottomrule
\end{tabular}%
}
\caption{Generation success rate under noisy (SNR=-10) and clean conditions with different denoising strategies.}
\label{tab:SEEN_ASR_appen}
\end{table*}
\begin{table}[t]
\centering
\resizebox{\columnwidth}{!}{%
\begin{tabular}{@{}l|llll@{}}
\toprule
\textbf{Method} & \textbf{Gauss}   & \textbf{Crowd}   & \textbf{Traffic} & \textbf{Machine} \\ \midrule
\textbf{None}   & 74.78\%          & 69.67\%          & 67.89\%          & 73.45\%          \\
\rowcolor[HTML]{EFEFEF} 
\textbf{STFT}   & 68.00\%          & 66.00\%          & 68.44\%          & 71.22\%          \\
\textbf{WT} & 63.22\%          & 67.44\%          & 64.67\%          & 60.78\%          \\
\rowcolor[HTML]{EFEFEF} 
\textbf{Segan}  & 64.78\%          & 64.66\%          & 64.67\%          & 64.33\%          \\
\textbf{DFL}    & 64.22\%          & 64.11\%          & 63.00\%          & 64.22\%          \\
\rowcolor[HTML]{EFEFEF} 
\textbf{SEEN}   & \textbf{75.22\%} & \textbf{70.00\%} & \textbf{68.00\%} & \textbf{74.67\%} \\ \bottomrule
\end{tabular}%
}
\caption{SEEN improves generation quality across noise types.}
\label{tab:More_noise_ASR_appen}
\end{table}
Compared with raw noise corruption, SEEN improves task performance across multiple datasets, although the gains are limited, as shown in Table~\ref{tab:SEEN_ASR_appen} and Table~\ref{tab:More_noise_ASR_appen}.
Regarding this phenomenon, we have the following conjectures.
SEEN operates by suppressing activations within the noise-aligned subspace, and thus can only remove noise-related components from the representation. 
However, when noise corrupts the semantic content of the input itself, the resulting task information becomes incomplete or ambiguous, which cannot be recovered through representation-level suppression alone. 
Consequently, if a model lacks robustness to incomplete or degraded inputs, applying SEEN in isolation is unlikely to yield substantial performance improvements.
Despite this constraint, SEEN consistently yields modest improvements, indicating that reducing noise-induced bias provides measurable benefits even when information loss dominates. 


\begin{table}[htbp]
\centering
\resizebox{\columnwidth}{!}{%
\begin{tabular}{@{}lcccc@{}}
\toprule
\textbf{Method} & \textbf{STFT} & \textbf{WT} & \textbf{Segan} & \textbf{DFL} \\
\midrule
\textbf{Time (s)}   & 23.11 & 2.21 & 4.67 & 1156.15 \\
\textbf{Complexity} & $O(N \log N)$ & $O(N)$ & $O(N \cdot L \cdot C)$ & $O(N \cdot L \cdot K^2)$ \\
\bottomrule
\end{tabular}%
}
\caption{Efficiency Analysis: Execution Time and Time Complexity for 100 Audio Samples}
\label{tab:shiyan}
\end{table}

\section{The transferability of SEE  metric}
\label{Sec:cross-over}
\begin{table*}[t]
\centering
\resizebox{\textwidth}{!}{%
\begin{tabular}{@{}l|l|cccc|cccc|cccc@{}}
\toprule
\multirow{2}{*}{\textbf{Dataset}} & \multirow{2}{*}{\textbf{SNR}} & \multicolumn{4}{c|}{\textbf{MiniCPM}} & \multicolumn{4}{c|}{\textbf{Qwen}} & \multicolumn{4}{c}{\textbf{StepAudio}} \\ \cmidrule(l){3-14} 
 &  & \textbf{Gauss} & \textbf{Crowd} & \textbf{Traffic} & \textbf{Machine} & \textbf{Gauss} & \textbf{Crowd} & \textbf{Traffic} & \textbf{Machine} & \textbf{Gauss} & \textbf{Crowd} & \textbf{Traffic} & \textbf{Machine} \\ \midrule
\multirow{4}{*}{\textbf{Music}} & \textbf{Clean} & 3.136 & 3.136 & 3.136 & 3.136 & 0.891 & 0.891 & 0.891 & 0.891 & 5.430 & 5.430 & 5.430 & 5.430 \\
 & \textbf{10} & 3.016 & 3.290 & 3.140 & 3.323 & 1.578 & 1.498 & 1.414 & 1.868 & 7.623 & 6.780 & 6.884 & 7.132 \\
 & \textbf{0} & 3.228 & 3.579 & 3.369 & 3.440 & 1.895 & 2.217 & 2.283 & 2.643 & 8.163 & 8.458 & 8.310 & 8.370 \\
 & \textbf{-10} & 3.567 & 3.569 & 3.562 & 3.455 & 2.453 & 3.159 & 3.616 & 3.27 & 8.612 & 9.654 & 9.291 & 9.770 \\ \midrule
\multirow{4}{*}{\textbf{Sound}} & \textbf{Clean} & 3.145 & 3.145 & 3.145 & 3.145 & 1.533 & 1.533 & 1.533 & 1.533 & 7.884 & 7.884 & 7.884 & 7.884 \\
 & \textbf{10} & 3.060 & 3.161 & 3.056 & 3.235 & 1.956 & 1.919 & 1.853 & 2.344 & 8.416 & 8.337 & 8.426 & 8.363 \\
 & \textbf{0} & 3.223 & 3.377 & 3.279 & 3.308 & 2.209 & 2.358 & 2.365 & 2.972 & 8.449 & 8.603 & 8.516 & 9.025 \\
 & \textbf{-10} & 3.549 & 3.545 & 3.634 & 3.350 & 2.420 & 3.044 & 3.346 & 3.342 & 8.168 & 8.967 & 8.351 & 9.906 \\ \midrule
\multirow{4}{*}{\textbf{Speech}} & \textbf{Clean} & 2.461 & 2.461 & 2.461 & 2.461 & 1.839 & 1.839 & 1.839 & 1.839 & 7.322 & 7.322 & 7.322 & 7.322 \\
 & \textbf{10} & 2.764 & 2.747 & 2.641 & 2.871 & 2.007 & 2.026 & 1.945 & 2.028 & 7.749 & 8.014 & 7.859 & 7.834 \\
 & \textbf{0} & 3.085 & 3.067 & 2.985 & 3.089 & 1.798 & 2.034 & 1.904 & 2.093 & 8.453 & 9.004 & 8.309 & 8.194 \\
 & \textbf{-10} & 3.703 & 3.706 & 3.862 & 3.391 & 1.768 & 2.353 & 2.681 & 2.354 & 8.583 & 9.703 & 8.890 & 8.559 \\ \bottomrule
\end{tabular}%
}
\caption{Cross-Noise Transferability of the SEE Metric across Different Noise Conditions}
\label{tab:cross_noise_SEE}
\end{table*}


To evaluate the transferability and generality of the proposed SEE metric, we conducted extensive cross-model evaluations across three distinct ALLMs: MiniCPM, Qwen, and StepAudio. The experimental protocol involved extracting directional vectors using a limited calibration set of 50 samples per noise category. These vectors were subsequently applied to a larger, unseen test suite comprising 300 samples per noise type.

The experimental results, summarized in table ~\ref{tab:cross_noise_SEE}, consistently demonstrate that across all three models and various acoustic environments (Gauss, Crowd, Traffic, and Machine), the SEE metric exhibits a stable monotonic relationship with the Signal-to-Noise Ratio (SNR). Specifically, as the SNR decreases from 10dB to -10dB, the SEE values increase progressively. This consistent trend across diverse model architectures and noise conditions underscores the strong transferability and reliability of the SEE metric in characterizing audio degradation. 

\section{Comparison of Representation Similarity across Enhancement Methods}
\label{app:sim}
\begin{table*}[t]
\centering
\resizebox{\textwidth}{!}{%
\begin{tabular}{@{}c|l|cc|cc|cc@{}}
\toprule
\multicolumn{1}{l|}{\multirow{2}{*}{\textbf{Dataset}}} & \multirow{2}{*}{\textbf{Method}} & \multicolumn{2}{c|}{\textbf{MiniCPM}} & \multicolumn{2}{c|}{\textbf{Qwen}} & \multicolumn{2}{c}{\textbf{StepAudio}} \\ \cmidrule(l){3-8} 
\multicolumn{1}{l|}{} &  & \textbf{Clean\_vs\_Noisy} & \textbf{Clean\_vs\_Enhanced} & \textbf{Clean\_vs\_Noisy} & \textbf{Clean\_vs\_Enhanced} & \textbf{Clean\_vs\_Noisy} & \textbf{Clean\_vs\_Enhanced} \\ \midrule
\multirow{5}{*}{\textbf{Music}} & \textbf{STFT} & 0.7138 & 0.6121 & 0.7549 & 0.7048 & 0.7121 & 0.6897 \\
 & \textbf{WT} & 0.7138 & 0.6108 & 0.7549 & 0.6335 & 0.7121 & 0.6543 \\
 & \textbf{Segan} & 0.7138 & 0.6273 & 0.7549 & 0.6304 & 0.7121 & 0.6481 \\
 & \textbf{DFL} & 0.7138 & 0.6775 & 0.7549 & 0.6944 & 0.7121 & 0.6736 \\
 & \textbf{SEEN} & \textbf{0.7138} & \textbf{0.7341} & \textbf{0.7549} & \textbf{0.7628} & \textbf{0.7121} & \textbf{0.7142} \\ \midrule
\multirow{5}{*}{\textbf{Sound}} & \textbf{STFT} & 0.7446 & 0.6339 & 0.7754 & 0.6938 & 0.8627 & 0.7924 \\
 & \textbf{WT} & 0.7446 & 0.6423 & 0.7754 & 0.6410 & 0.8627 & 0.7263 \\
 & \textbf{Segan} & 0.7446 & 0.6880 & 0.7754 & 0.7098 & 0.8627 & 0.8015 \\
 & \textbf{DFL} & 0.7446 & 0.7284 & 0.7754 & 0.6880 & 0.8627 & 0.7804 \\
 & \textbf{SEEN} & \textbf{0.7446} & \textbf{0.7858} & \textbf{0.7754} & \textbf{0.7827} & \textbf{0.8627} & \textbf{0.8631} \\ \midrule
\multirow{5}{*}{\textbf{Speech}} & \textbf{STFT} & 0.5837 & 0.4708 & 0.7271 & 0.6880 & 0.8399 & 0.8119 \\
 & \textbf{WT} & 0.5837 & 0.3683 & 0.7271 & 0.5345 & 0.8399 & 0.6943 \\
 & \textbf{Segan} & 0.5837 & 0.3997 & 0.7271 & 0.6303 & 0.8399 & 0.7831 \\
 & \textbf{DFL} & 0.5837 & 0.5439 & 0.7271 & 0.6588 & 0.8399 & 0.7986 \\
 & \textbf{SEEN} & \textbf{0.5837} & \textbf{0.6579} & \textbf{0.7271} & \textbf{0.7345} & \textbf{0.8399} & \textbf{0.8476} \\ \bottomrule
\end{tabular}%
}
\caption{Cosine Similarity Between Clean Speech and Noisy or Enhanced Speech Under Different Enhancement Methods and Noise Conditions(SNR=-5)}
\label{tab:sim--5}
\end{table*}
\begin{table*}[t]
\centering
\resizebox{\textwidth}{!}{%
\begin{tabular}{@{}c|l|cc|cc|cc@{}}
\toprule
\multicolumn{1}{l|}{\multirow{2}{*}{\textbf{Dataset}}} & \multirow{2}{*}{\textbf{Method}} & \multicolumn{2}{c|}{\textbf{MiniCPM}} & \multicolumn{2}{c|}{\textbf{Qwen}} & \multicolumn{2}{c}{\textbf{StepAudio}} \\ \cmidrule(l){3-8} 
\multicolumn{1}{l|}{} &  & \textbf{Clean\_vs\_Noisy} & \textbf{Clean\_vs\_Enhanced} & \textbf{Clean\_vs\_Noisy} & \textbf{Clean\_vs\_Enhanced} & \textbf{Clean\_vs\_Noisy} & \textbf{Clean\_vs\_Enhanced} \\ \midrule
\multirow{5}{*}{\textbf{Music}} & \textbf{STFT} & 0.6523 & 0.5965 & 0.6701 & 0.6551 & 0.7031 & 0.6912 \\
 & \textbf{WT} & 0.6523 & 0.5752 & 0.6701 & 0.6390 & 0.7031 & 0.6566 \\
 & \textbf{Segan} & 0.6523 & 0.6078 & 0.6701 & 0.6012 & 0.7031 & 0.6459 \\
 & \textbf{DFL} & 0.6523 & 0.6525 & 0.6701 & 0.6648 & 0.7031 & 0.6656 \\
 & \textbf{SEEN} & \textbf{0.6523} & \textbf{0.6509} & \textbf{0.6701} & \textbf{0.6823} & \textbf{0.7031} & \textbf{0.7002} \\ \midrule
\multirow{5}{*}{\textbf{Sound}} & \textbf{STFT} & 0.7129 & 0.6188 & 0.7034 & 0.6460 & 0.8291 & 0.7641 \\
 & \textbf{WT} & 0.7129 & 0.6031 & 0.7034 & 0.6297 & 0.8291 & 0.6921 \\
 & \textbf{Segan} & 0.7129 & 0.6693 & 0.7034 & 0.6557 & 0.8291 & 0.7713 \\
 & \textbf{DFL} & 0.7129 & 0.6951 & 0.7034 & 0.6512 & 0.8291 & 0.7510 \\
 & \textbf{SEEN} & \textbf{0.7129} & \textbf{0.7124} & \textbf{0.7034} & \textbf{0.7125} & \textbf{0.8291} & \textbf{0.8114} \\ \midrule
\multirow{5}{*}{\textbf{Speech}} & \textbf{STFT} & 0.5314 & 0.4141 & 0.6452 & 0.6058 & 0.8060 & 0.7692 \\
 & \textbf{WT} & 0.5314 & 0.3365 & 0.6452 & 0.4967 & 0.8069 & 0.6932 \\
 & \textbf{Segan} & 0.5314 & 0.3760 & 0.6452 & 0.5359 & 0.8060 & 0.7410 \\
 & \textbf{DFL} & 0.5314 & 0.4517 & 0.6452 & 0.5867 & 0.8060 & 0.7560 \\
 & \textbf{SEEN} & \textbf{0.5314} & \textbf{0.5292} & \textbf{0.6452} & \textbf{0.6474} & \textbf{0.8069} & \textbf{0.7868} \\ \bottomrule
\end{tabular}%
}
\caption{Cosine Similarity Between Clean Speech and Noisy or Enhanced Speech Under Different Enhancement Methods and Noise Conditions(SNR=-10)}
\label{tab:sim--10}
\end{table*}
\begin{table*}[t]
\centering
\resizebox{\textwidth}{!}{%
\begin{tabular}{@{}c|l|cc|cc|cc@{}}
\toprule
\multicolumn{1}{l|}{\multirow{2}{*}{\textbf{Dataset}}} & \multirow{2}{*}{\textbf{Method}} & \multicolumn{2}{c|}{\textbf{MiniCPM}} & \multicolumn{2}{c|}{\textbf{Qwen}} & \multicolumn{2}{c}{\textbf{StepAudio}} \\ \cmidrule(l){3-8} 
\multicolumn{1}{l|}{} &  & \textbf{Clean\_vs\_Noisy} & \textbf{Clean\_vs\_Enhanced} & \textbf{Clean\_vs\_Noisy} & \textbf{Clean\_vs\_Enhanced} & \textbf{Clean\_vs\_Noisy} & \textbf{Clean\_vs\_Enhanced} \\ \midrule
\multirow{5}{*}{\textbf{Music}} & \textbf{STFT} & 0.7967 & 0.6714 & 0.8280 & 0.7631 & 0.7435 & 0.6994 \\
 & \textbf{WT} & 0.7967 & 0.6394 & 0.8280 & 0.6539 & 0.7435 & 0.6553 \\
 & \textbf{Segan} & 0.7967 & 0.6861 & 0.8280 & 0.6895 & 0.7435 & 0.6629 \\
 & \textbf{DFL} & 0.7967 & 0.7084 & 0.8280 & 0.7275 & 0.7435 & 0.6847 \\
 & \textbf{SEEN} & \textbf{0.7967} & \textbf{0.8099} & \textbf{0.8280} & \textbf{0.8312} & \textbf{0.7435} & \textbf{0.7371} \\ \midrule
\multirow{5}{*}{\textbf{Sound}} & \textbf{STFT} & 0.8129 & 0.6849 & 0.8418 & 0.7420 & 0.8928 & 0.8170 \\
 & \textbf{WT} & 0.8129 & 0.6762 & 0.8418 & 0.6675 & 0.8928 & 0.7460 \\
 & \textbf{Segan} & 0.8129 & 0.7488 & 0.8418 & 0.7669 & 0.8928 & 0.8267 \\
 & \textbf{DFL} & 0.8129 & 0.7609 & 0.8418 & 0.7235 & 0.8928 & 0.8068 \\
 & \textbf{SEEN} & \textbf{0.8129} & \textbf{0.8486} & \textbf{0.8180} & \textbf{0.8462} & \textbf{0.8928} & \textbf{0.8718} \\ \midrule
\multirow{5}{*}{\textbf{Speech}} & \textbf{STFT} & 0.7042 & 0.6155 & 0.8097 & 0.7703 & 0.8495 & 0.8446 \\
 & \textbf{WT} & 0.7042 & 0.4365 & 0.8097 & 0.6083 & 0.8495 & 0.7077 \\
 & \textbf{Segan} & 0.7042 & 0.5320 & 0.8097 & 0.7206 & 0.8495 & 0.8186 \\
 & \textbf{DFL} & 0.7042 & 0.6295 & 0.8097 & 0.7253 & 0.8495 & 0.8348 \\
 & \textbf{SEEN} & \textbf{0.7042} & \textbf{0.7257} & \textbf{0.8097} & \textbf{0.8082} & \textbf{0.8495} & \textbf{0.8486} \\ \bottomrule
\end{tabular}%
}
\caption{Cosine Similarity Between Clean Speech and Noisy or Enhanced Speech Under Different Enhancement Methods and Noise Conditions(SNR=0)}
\label{tab:sim-0}
\end{table*}
\begin{table*}[t]
\centering
\resizebox{\textwidth}{!}{%
\begin{tabular}{@{}c|l|cc|cc|cc@{}}
\toprule
\multicolumn{1}{l|}{\multirow{2}{*}{\textbf{Dataset}}} & \multirow{2}{*}{\textbf{Method}} & \multicolumn{2}{c|}{\textbf{MiniCPM}} & \multicolumn{2}{c|}{\textbf{Qwen}} & \multicolumn{2}{c}{\textbf{StepAudio}} \\ \cmidrule(l){3-8} 
\multicolumn{1}{l|}{} &  & \textbf{Clean\_vs\_Noisy} & \textbf{Clean\_vs\_Enhanced} & \textbf{Clean\_vs\_Noisy} & \textbf{Clean\_vs\_Enhanced} & \textbf{Clean\_vs\_Noisy} & \textbf{Clean\_vs\_Enhanced} \\ \midrule
\multirow{5}{*}{\textbf{Music}} & \textbf{STFT} & 0.8566 & 0.7610 & 0.8750 & 0.8083 & 0.7702 & 0.7220 \\
 & \textbf{WT} & 0.8566 & 0.6815 & 0.8750 & 0.6955 & 0.7702 & 0.6724 \\
 & \textbf{Segan} & 0.8566 & 0.7659 & 0.8750 & 0.7599 & 0.7702 & 0.6887 \\
 & \textbf{DFL} & 0.8566 & 0.7326 & 0.8750 & 0.7562 & 0.7702 & 0.6880 \\
 & \textbf{SEEN} & \textbf{0.8566} & \textbf{0.8636} & \textbf{0.8750} & \textbf{0.8749} & \textbf{0.7702} & \textbf{0.7822} \\ \midrule
\multirow{5}{*}{\textbf{Sound}} & \textbf{STFT} & 0.8681 & 0.7491 & 0.8909 & 0.7817 & 0.9184 & 0.8390 \\
 & \textbf{WT} & 0.8681 & 0.7268 & 0.8909 & 0.7104 & 0.9184 & 0.7899 \\
 & \textbf{Segan} & 0.8681 & 0.8075 & 0.8909 & 0.8134 & 0.9184 & 0.8455 \\
 & \textbf{DFL} & 0.8681 & 0.7864 & 0.8909 & 0.7556 & 0.9184 & 0.8290 \\
 & \textbf{SEEN} & \textbf{0.8681} & \textbf{0.8948} & \textbf{0.8909} & \textbf{0.8923} & \textbf{0.9184} & \textbf{0.9067} \\ \midrule
\multirow{5}{*}{\textbf{Speech}} & \textbf{STFT} & 0.7990 & 0.7449 & 0.8683 & 0.8297 & 0.9002 & 0.8774 \\
 & \textbf{WT} & 0.7990 & 0.5480 & 0.8683 & 0.6902 & 0.9002 & 0.7549 \\
 & \textbf{Segan} & 0.7990 & 0.6618 & 0.8683 & 0.8002 & 0.9002 & 0.8392 \\
 & \textbf{DFL} & 0.7990 & 0.7105 & 0.8683 & 0.7909 & 0.9002 & 0.8617 \\
 & \textbf{SEEN} & \textbf{0.7990} & \textbf{0.8213} & \textbf{0.8683} & \textbf{0.8757} & \textbf{0.9002} & \textbf{0.8944} \\ \bottomrule
\end{tabular}%
}
\caption{Cosine Similarity Between Clean Speech and Noisy or Enhanced Speech Under Different Enhancement Methods and Noise Conditions(SNR=5)}
\label{tab:sim-5}
\end{table*}
\begin{table*}[t]
\centering
\resizebox{\textwidth}{!}{%
\begin{tabular}{@{}c|l|cc|cc|cc@{}}
\toprule
\multicolumn{1}{l|}{\multirow{2}{*}{\textbf{Dataset}}} & \multirow{2}{*}{\textbf{Method}} & \multicolumn{2}{c|}{\textbf{MiniCPM}} & \multicolumn{2}{c|}{\textbf{Qwen}} & \multicolumn{2}{c}{\textbf{StepAudio}} \\ \cmidrule(l){3-8} 
\multicolumn{1}{l|}{} &  & \textbf{Clean\_vs\_Noisy} & \textbf{Clean\_vs\_Enhanced} & \textbf{Clean\_vs\_Noisy} & \textbf{Clean\_vs\_Enhanced} & \textbf{Clean\_vs\_Noisy} & \textbf{Clean\_vs\_Enhanced} \\ \midrule
\multirow{5}{*}{\textbf{Music}} & \textbf{STFT} & 0.8979 & 0.8319 & 0.9066 & 0.8404 & 0.7984 & 0.7468 \\
 & \textbf{WT} & 0.8979 & 0.7395 & 0.9066 & 0.7493 & 0.7984 & 0.6942 \\
 & \textbf{Segan} & 0.8979 & 0.8209 & 0.9066 & 0.8200 & 0.7984 & 0.7153 \\
 & \textbf{DFL} & 0.8979 & 0.7522 & 0.9066 & 0.7782 & 0.7984 & 0.6917 \\
 & \textbf{SEEN} & \textbf{0.8979} & \textbf{0.9035} & \textbf{0.9066} & \textbf{0.9048} & \textbf{0.7984} & \textbf{0.7891} \\ \midrule
\multirow{5}{*}{\textbf{Sound}} & \textbf{STFT} & 0.9076 & 0.7966 & 0.9243 & 0.8095 & 0.9393 & 0.8495 \\
 & \textbf{WT} & 0.9076 & 0.7864 & 0.9243 & 0.7646 & 0.9393 & 0.8384 \\
 & \textbf{Segan} & 0.9076 & 0.8489 & 0.9243 & 0.8442 & 0.9393 & 0.8534 \\
 & \textbf{DFL} & 0.9076 & 0.8032 & 0.9243 & 0.7779 & 0.9393 & 0.8445 \\
 & \textbf{SEEN} & \textbf{0.9076} & \textbf{0.9274} & \textbf{0.9243} & \textbf{0.9234} & \textbf{0.9303} & \textbf{0.9172} \\ \midrule
\multirow{5}{*}{\textbf{Speech}} & \textbf{STFT} & 0.8635 & 0.8304 & 0.9018 & 0.8615 & 0.9221 & 0.9010 \\
 & \textbf{WT} & 0.8635 & 0.6704 & 0.9018 & 0.7695 & 0.9221 & 0.8111 \\
 & \textbf{Segan} & 0.8635 & 0.7550 & 0.9018 & 0.8514 & 0.9221 & 0.8613 \\
 & \textbf{DFL} & 0.8635 & 0.7745 & 0.9018 & 0.8360 & 0.9221 & 0.8801 \\
 & \textbf{SEEN} & \textbf{0.8635} & \textbf{0.8893} & \textbf{0.9018} & \textbf{0.8983} & \textbf{0.9221} & \textbf{0.9050} \\ \bottomrule
\end{tabular}%
}
\caption{Cosine Similarity Between Clean Speech and Noisy or Enhanced Speech Under Different Enhancement Methods and Noise Conditions(SNR=10)}
\label{tab:sim-10}
\end{table*}
\begin{table*}[t]
\centering
\resizebox{\textwidth}{!}{%
\begin{tabular}{@{}c|l|cc|cc|cc@{}}
\toprule
\multicolumn{1}{l|}{\multirow{2}{*}{\textbf{Dataset}}} & \multirow{2}{*}{\textbf{Method}} & \multicolumn{2}{c|}{\textbf{MiniCPM}} & \multicolumn{2}{c|}{\textbf{Qwen}} & \multicolumn{2}{c}{\textbf{StepAudio}} \\ \cmidrule(l){3-8} 
\multicolumn{1}{l|}{} &  & \textbf{Clean\_vs\_Noisy} & \textbf{Clean\_vs\_Enhanced} & \textbf{Clean\_vs\_Noisy} & \textbf{Clean\_vs\_Enhanced} & \textbf{Clean\_vs\_Noisy} & \textbf{Clean\_vs\_Enhanced} \\ \midrule
\multirow{5}{*}{\textbf{Music}} & \textbf{STFT} & 0.9530 & 0.8985 & 0.9444 & 0.8886 & 0.8480 & 0.7906 \\
 & \textbf{WT} & 0.9530 & 0.8340 & 0.9444 & 0.8427 & 0.8480 & 0.7551 \\
 & \textbf{Segan} & 0.9530 & 0.8854 & 0.9444 & 0.8749 & 0.8480 & 0.7570 \\
 & \textbf{DFL} & 0.9530 & 0.7716 & 0.9444 & 0.7988 & 0.8480 & 0.7039 \\
 & \textbf{SEEN} & \textbf{0.9530} & \textbf{0.9492} & \textbf{0.9444} & \textbf{0.9413} & \textbf{0.8480} & \textbf{0.8361} \\ \midrule
\multirow{5}{*}{\textbf{Sound}} & \textbf{STFT} & 0.9687 & 0.8667 & 0.9621 & 0.8483 & 0.9661 & 0.8804 \\
 & \textbf{WT} & 0.9687 & 0.8819 & 0.9621 & 0.8596 & 0.9661 & 0.9046 \\
 & \textbf{Segan} & 0.9687 & 0.8978 & 0.9621 & 0.8719 & 0.9661 & 0.8618 \\
 & \textbf{DFL} & 0.9687 & 0.8215 & 0.9621 & 0.8004 & 0.9661 & 0.8588 \\
 & \textbf{SEEN} & \textbf{0.9687} & \textbf{0.9650} & \textbf{0.9621} & \textbf{0.9590} & \textbf{0.9661} & \textbf{0.9436} \\ \midrule
\multirow{5}{*}{\textbf{Speech}} & \textbf{STFT} & 0.9404 & 0.9186 & 0.9345 & 0.9016 & 0.9590 & 0.9292 \\
 & \textbf{WT} & 0.9404 & 0.8445 & 0.9345 & 0.8928 & 0.9590 & 0.8782 \\
 & \textbf{Segan} & 0.9404 & 0.8698 & 0.9345 & 0.8863 & 0.9590 & 0.8784 \\
 & \textbf{DFL} & 0.9404 & 0.8391 & 0.9345 & 0.8699 & 0.9590 & 0.9025 \\
 & \textbf{SEEN} & \textbf{0.9404} & \textbf{0.9379} & \textbf{0.9343} & \textbf{0.9307} & \textbf{0.9590} & \textbf{0.9497} \\ \bottomrule
\end{tabular}%
}
\caption{Cosine Similarity Between Clean Speech and Noisy or Enhanced Speech Under Different Enhancement Methods and Noise Conditions(SNR=20)}
\label{tab:sim-20}
\end{table*}
\begin{table*}[t]
\centering
\resizebox{\textwidth}{!}{%
\begin{tabular}{@{}c|l|cc|cc|cc@{}}
\toprule
\multicolumn{1}{l|}{\multirow{2}{*}{\textbf{Dataset}}} & \multirow{2}{*}{\textbf{Method}} & \multicolumn{2}{c|}{\textbf{MiniCPM}} & \multicolumn{2}{c|}{\textbf{Qwen}} & \multicolumn{2}{c}{\textbf{StepAudio}} \\ \cmidrule(l){3-8} 
\multicolumn{1}{l|}{} &  & \textbf{Clean\_vs\_Noisy} & \textbf{Clean\_vs\_Enhanced} & \textbf{Clean\_vs\_Noisy} & \textbf{Clean\_vs\_Enhanced} & \textbf{Clean\_vs\_Noisy} & \textbf{Clean\_vs\_Enhanced} \\ \midrule
\multirow{5}{*}{\textbf{Music}} & \textbf{STFT} & 0.9783 & 0.9137 & 0.9671 & 0.9092 & 0.8967 & 0.8196 \\
 & \textbf{WT} & 0.9783 & 0.8818 & 0.9671 & 0.8892 & 0.8967 & 0.7968 \\
 & \textbf{Segan} & 0.9783 & 0.9070 & 0.9671 & 0.8938 & 0.8967 & 0.7832 \\
 & \textbf{DFL} & 0.9783 & 0.7840 & 0.9671 & 0.8121 & 0.8967 & 0.7109 \\
 & \textbf{SEEN} & \textbf{0.9783} & \textbf{0.9743} & \textbf{0.9671} & \textbf{0.9637} & \textbf{0.8967} & \textbf{0.8822} \\ \midrule
\multirow{5}{*}{\textbf{Sound}} & \textbf{STFT} & 0.9873 & 0.8835 & 0.9811 & 0.8672 & 0.9809 & 0.8960 \\
 & \textbf{WT} & 0.9873 & 0.9204 & 0.9811 & 0.9027 & 0.9809 & 0.8912 \\
 & \textbf{Segan} & 0.9873 & 0.9092 & 0.9811 & 0.8827 & 0.9809 & 0.8637 \\
 & \textbf{DFL} & 0.9873 & 0.8320 & 0.9811 & 0.8123 & 0.9809 & 0.8670 \\
 & \textbf{SEEN} & \textbf{0.9873} & \textbf{0.9833} & \textbf{0.9811} & \textbf{0.9772} & \textbf{0.9809} & \textbf{0.9582} \\ \midrule
\multirow{5}{*}{\textbf{Speech}} & \textbf{STFT} & 0.9605 & 0.9371 & 0.9527 & \textbf{0.9253} & 0.9816 & 0.9418 \\
 & \textbf{WT} & 0.9605 & 0.9186 & 0.9527 & 0.9403 & 0.9816 & 0.9038 \\
 & \textbf{Segan} & 0.9605 & 0.8880 & 0.9527 & 0.8946 & 0.9816 & 0.8833 \\
 & \textbf{DFL} & 0.9605 & 0.8726 & 0.9527 & 0.8878 & 0.9816 & 0.9173 \\
 & \textbf{SEEN} & \textbf{0.9605} & \textbf{0.9580} & \textbf{0.9527} & \textbf{0.9588} & \textbf{0.9816} & \textbf{0.9612} \\ \bottomrule
\end{tabular}%
}
\caption{Cosine Similarity Between Clean Speech and Noisy or Enhanced Speech Under Different Enhancement Methods and Noise Conditions(SNR=30)}
\label{tab:sim-30}
\end{table*}
Tables~\ref{tab:sim--5}--\ref{tab:sim-30} reports the cosine similarity between clean speech and its noisy or enhanced counterparts under different noise levels and enhancement methods. Specifically, \texttt{Clean_vs_Noisy} measures the similarity between clean and noisy speech, while \texttt{Clean_vs_Enhanced} measures the similarity between clean speech and speech processed by different enhancement methods.

As shown in the results, conventional speech enhancement methods often fail to reduce the representation gap between clean and noisy speech. In many cases, they even enlarge this gap, resulting in lower cosine similarity compared to the original noisy input. In contrast, our proposed method consistently narrows the similarity gap between clean and noisy speech, indicating its superior ability to preserve semantic and acoustic consistency under noise perturbations. This auxiliary experiment provides additional evidence supporting the superiority of our proposed method over traditional approaches.

\section{Efficiency Analysis About Baseline Methods}

To evaluate computational efficiency, we conducted a benchmark on 100 audio samples from the Speech dataset with SNR=0 dB. As shown in Table~\ref{tab:shiyan}, external denoising methods inevitably introduce additional processing time: the WT and STFT methods require 2.21s and 23.11s respectively, while the DFL model reaches 1156.15s. Owing to the limited CUDA version support of the DFL implementation, we performed the evaluation on the CPU. While applying traditional or external deep learning denoising to ALLMs typically incurs such "bottleneck" delays, our method introduces zero additional processing time, enabling more seamless integration for real-time applications.

The time complexity of the evaluated methods in inference phase is also summarized in Table \ref{tab:shiyan}. Here, $N$ represents the total number of audio samples in a signal. For the deep learning models (Segan and DFL), $L$ denotes the number of hidden layers, $C$ represents the number of feature channels, and $K$ indicates the size of the convolutional kernels.

\section{Validity Statement of SEEN}
\label{Sec:SEEN_validity_statement}
In our experiments, SEE consistently increases under stronger corruption and remains well separated from the clean-input range, indicating that it captures a stable noise-aligned component in the embedding space. However, applying SEEN to suppress this component yields only a modest but consistent improvement in generation quality, rather than a large robustness recovery. This gap between measurability and recoverability suggests that the interference identified by SEE is one of core component of the mechanism driving downstream failures.

A plausible explanation is that the subspace isolated by SEE represents directions along which noise reliably perturbs the model's internal coding tendency, so removing these directions can reduce embedding bias and mitigate spurious activation. Yet generation errors are not solely caused by biased representations; when noise is strong, it can also erase or distort task-relevant acoustic cues at the source, producing genuine information loss that cannot be recovered by subtracting a projection term. In such cases, SEEN can remove interference but cannot reconstruct missing content, so the attainable improvement is bounded by the amount of recoverable information rather than by the clean–noisy performance gap. This effect may be particularly pronounced for audio inputs. Frame-level tokenization typically produces long token sequences with relatively low per-token information density, meaning that corruption can distribute small distortions across many tokens and accumulate into semantic degradation that is difficult to reverse through a single subspace suppression.

These observations point to a natural direction for further improvement. While SEE provides a useful criterion for diagnosing and comparing noise-induced interference, stronger robustness may require training-time mechanisms that explicitly enhance the model's ability to infer semantics under partial or degraded acoustic evidence. For example, one could consider incorporating SEE-guided objectives into robustness-oriented training, or adopting augmentation curricula that emphasize semantic preservation under frame-level corruption. We view these as promising hypotheses rather than confirmed solutions, and leave a systematic exploration of training-side enhancements as an important next step.

\section{Algorithm Flow}
In this appendix, we present the formal procedures for our proposed framework SEEN. The process is divided into three main components: algorithm 1: layer localization and noise extraction (\autoref{alg:see_setup}), algorithm 2: energy-based evaluation (\autoref{alg:see_eval}), and algorithm 3: the final robust inference (\autoref{alg:seen_infer}). Together, these algorithms form a complete pipeline for diagnosing and neutralizing noise-aligned activations in Large Audio Language Models.

\section{AI Writing Assistance Disclosure}
We used AI tools solely for language polishing to improve clarity and readability. The AI tools did not contribute to the scientific content, ideas, analyses, or conclusions of this work.

\begin{algorithm*}[t]
\caption{SEE Setup: Layer Localization and Noise Extraction}
\label{alg:see_setup}
\begin{algorithmic}[1]
\REQUIRE Semantic set $X^s=\{x_i^s\}_{i=1}^{m}$, noise set $X^n=\{x_i^n\}_{i=1}^{m}$, thresholds $\alpha,\delta$, stabilizer $\varepsilon$
\ENSURE Retained layers $\mathcal{L}^\star$, noise bases $\{\mathbf{Q}_\ell\}_{\ell\in\mathcal{L}^\star}$

\FOR{$\ell=1$ \TO $L$}
    \FOR{$i=1$ \TO $m$}
        \STATE Extract $\mathbf{A}_\ell(x_i^s)\in\mathbb{R}^{T(x_i^s)\times d_\ell}$ and $\mathbf{A}_\ell(x_i^n)\in\mathbb{R}^{T(x_i^n)\times d_\ell}$
        \STATE $\mathbf{a}_\ell(x_i^s)\leftarrow \frac{1}{T(x_i^s)}\sum_{t=1}^{T(x_i^s)}\mathbf{A}_\ell(x_i^s)_{t,:}\ \in\mathbb{R}^{d_\ell}$
        \STATE $\mathbf{a}_\ell(x_i^n)\leftarrow \frac{1}{T(x_i^n)}\sum_{t=1}^{T(x_i^n)}\mathbf{A}_\ell(x_i^n)_{t,:}\ \in\mathbb{R}^{d_\ell}$
    \ENDFOR
    \STATE Stack $\{\mathbf{a}_\ell(x_i^s)\}_{i=1}^{m}$ to form $\mathbf{S}_\ell\in\mathbb{R}^{m\times d_\ell}$; stack $\{\mathbf{a}_\ell(x_i^n)\}_{i=1}^{m}$ to form $\mathbf{N}_\ell\in\mathbb{R}^{m\times d_\ell}$
    \STATE Magnitude: $\mathrm{M}_\ell \leftarrow \|\mathbf{S}_\ell-\mathbf{N}_\ell\|_F$
    \STATE Direction: $\mathrm{D}_\ell \leftarrow \frac{\left|\mathrm{vec}(\mathbf{S}_\ell)^\top \mathrm{vec}(\mathbf{N}_\ell)\right|}{\|\mathrm{vec}(\mathbf{S}_\ell)\|_2\,\|\mathrm{vec}(\mathbf{N}_\ell)\|_2+\varepsilon}$
\ENDFOR

\STATE $\overline{\mathrm{M}}\leftarrow \frac{1}{L}\sum_{\ell=1}^{L}\mathrm{M}_\ell$, \quad $\overline{\mathrm{D}}\leftarrow \frac{1}{L}\sum_{\ell=1}^{L}\mathrm{D}_\ell$
\STATE $\ell^\star \leftarrow \min\{\ell \mid \mathrm{M}_\ell>\overline{\mathrm{M}} \ \land\  \mathrm{D}_\ell>\overline{\mathrm{D}}\}$
\STATE $\mathcal{L}^\star \leftarrow \{\ell^\star,\ell^\star+1,\dots,L\}$

\FOR{each $\ell\in\mathcal{L}^\star$}
    \STATE SVD: $\mathbf{S}_\ell=\mathbf{U}^s_\ell\mathbf{\Sigma}^s_\ell(\mathbf{V}^s_\ell)^\top$; \quad $\mathbf{N}_\ell=\mathbf{U}^n_\ell\mathbf{\Sigma}^n_\ell(\mathbf{V}^n_\ell)^\top$
    \STATE $\mathcal{I}^s_\ell\leftarrow \{j\mid \sigma^s_{\ell,j}>\alpha\}$;\quad $\mathcal{I}^n_\ell\leftarrow \{j\mid \sigma^n_{\ell,j}>\alpha\}$
    \FOR{each $j\in\mathcal{I}^n_\ell$}
        \STATE $\textbf{m}_{\ell,j}\leftarrow \max_{k\in\mathcal{I}^s_\ell}\left|\mathrm{cos}(\mathbf{v}^n_{\ell,j},\mathbf{v}^s_{\ell,k})\right|$
    \ENDFOR
    \STATE $\mathcal{J}_\ell\leftarrow \{j\in\mathcal{I}^n_\ell \mid \textbf{m}_{\ell,j}<\delta\}$; \quad $r_\ell\leftarrow |\mathcal{J}_\ell|$
    \STATE Build mask $s_\ell\in\{0,1\}^{d_\ell}$ with $s_\ell[j]=1$ iff $j\in\mathcal{J}_\ell$; \quad $\mathbf{M}_\ell\leftarrow \mathrm{diag}(s_\ell)$
    \STATE $\mathbf{Q}_\ell \leftarrow \mathrm{NonZeroCols}(\mathbf{V}^n_\ell\mathbf{M}_\ell)\in\mathbb{R}^{d_\ell\times r_\ell}$ \COMMENT{equivalently keep columns $\mathcal{J}_\ell$ of $\mathbf{V}^n_\ell$}
\ENDFOR
\end{algorithmic}
\end{algorithm*}

\begin{algorithm*}[t]
\caption{SEE Eval: Computing Signal Embedding Energy for an Input}
\label{alg:see_eval}
\begin{algorithmic}[1]
\REQUIRE Test input $x$, retained layers $\mathcal{L}^\star$, noise bases $\{\mathbf{Q}_\ell\}_{\ell\in\mathcal{L}^\star}$, stabilizer $\varepsilon$
\ENSURE $\mathrm{SEE}(x)$
\STATE Forward to obtain $\mathbf{A}_\ell(x)\in\mathbb{R}^{T(x)\times d_\ell}$ for all $\ell\in\mathcal{L}^\star$
\FOR{each $\ell\in\mathcal{L}^\star$}
    \STATE Project: $\mathbf{Z}_\ell(x)\leftarrow \mathbf{A}_\ell(x)\mathbf{Q}_\ell \in \mathbb{R}^{T(x)\times r_\ell}$
    \STATE $\mathrm{SEE}_\ell(x)\leftarrow \frac{1}{T(x)}\sum_{t=1}^{T(x)}\frac{\|\mathbf{Z}_\ell(x)_{t,:}\|_2^2}{\|\mathbf{A}_\ell(x)_{t,:}\|_2^2+\varepsilon}$
\ENDFOR
\STATE $\mathrm{SEE}(x)\leftarrow \frac{1}{|\mathcal{L}^\star|}\sum_{\ell\in\mathcal{L}^\star}\mathrm{SEE}_\ell(x)$
\end{algorithmic}
\end{algorithm*}

\begin{algorithm*}[t]
\caption{SEEN Inference: Neutralizing Noise-aligned Activations}
\label{alg:seen_infer}
\begin{algorithmic}[1]
\REQUIRE Test input $x$, retained layers $\mathcal{L}^\star$, noise bases $\{\mathbf{Q}_\ell\}_{\ell\in\mathcal{L}^\star}$, neutralization strengths $\{\lambda\}_{\ell\in\mathcal{L}^\star}$
\ENSURE Neutralized activations $\{\widetilde{\mathbf{A}}_\ell(x)\}$ 
\STATE Forward to obtain $\mathbf{A}_\ell(x)\in\mathbb{R}^{T(x)\times d_\ell}$ for all $\ell\in\mathcal{L}^\star$
\FOR{each $\ell\in\mathcal{L}^\star$}
    \STATE Reconstruct noise component: $\mathbf{C}_\ell(x)\leftarrow \mathbf{A}_\ell(x)\mathbf{Q}_\ell\mathbf{Q}_\ell^\top \in \mathbb{R}^{T(x)\times d_\ell}$
    \STATE Neutralize: $\widetilde{\mathbf{A}}_\ell(x)\leftarrow \mathbf{A}_\ell(x)-\lambda\,\mathbf{C}_\ell(x)$
\ENDFOR
\STATE Continue forward pass using $\widetilde{\mathbf{A}}_\ell(x)$ to obtain the final generation output
\end{algorithmic}
\end{algorithm*}

\end{document}